\documentclass[a4paper,11pt]{article}%\documentstyle[12pt]{article}
\usepackage{amssymb}
\author{\small Wen-Xiu Ma$^\dagger$ and Zixiang Zhou$^\ddagger$\\
\small $^\dagger$ Department of Mathematics, City University of Hong Kong, 
Hong Kong, Kowloon, China\\
\small $^\ddagger$ Institute of Mathematics, Fudan University, Shanghai 200433, China
}

\title {
\begin{flushright}\normalsize  \sf
           {\tt nlin.SI/0105061}, May 2001
            \end{flushright}
Binary Symmetry Constraints of ${\cal N}$-wave 
Interaction
Equations in $1+1$ and $2+1$ Dimensions} 

\date{\nonumber}
\def\diag{\hbox{\rm diag}}
\def\D{\displaystyle}

\font\scriptrm=cmr8

\begin{document}

\setlength{\baselineskip}{16pt}

\maketitle

\begin{abstract}

Binary symmetry constraints of the ${\cal N}$-wave interaction equations 
in $1+1$ and $2+1$ dimensions are proposed to reduce 
the ${\cal N}$-wave interaction equations
into finite-dimensional Liouville integrable systems. 
A new involutive and functionally independent system 
of polynomial functions is generated from an arbitrary order square
matrix Lax operator and 
used to show the Liouville integrability of the 
constrained flows of the ${\cal N}$-wave interaction equations. 
The constraints on the potentials resulting from 
the symmetry constraints 
give rise to involutive solutions to 
the ${\cal N}$-wave interaction equations,
and thus the integrability by quadratures are shown for 
the ${\cal N}$-wave interaction equations
by the constrained flows.

\end{abstract}

Running title: Symmetry Constraints of ${\cal N}$-wave 
Equations 

\newcommand{\Z}{\mathbb{Z}}
\newcommand{\C}{\mathbb{C}}
\newcommand{\R}{\mathbb{R}}

\def\llsum#1#2{\sum_{\scriptstyle l_1+\cdots+l_{#1}=#2\atop\scriptstyle
l_1,\cdots,l_{#1}\ge 0}c_1^{l_1}\cdots c_{#1}^{l_{#1}}}

\def\sj#1#2{s_{#1}(c_1,\cdots,c_{#2})}

\def \ba {\begin{array}}
\def \ea {\end{array}}
\def\bea{\begin{eqnarray}}
\def\eea{\end{eqnarray}}
\def \part {\partial }

\renewcommand{\theequation}{\arabic{section}.\arabic{equation}}

%\renewcommand{\theequation}{\thesection.\arabic{equation}}

%\newcounter{theorem1}{\arabic{section}.\arabic{theorem}}
\newtheorem{lemma}{Lemma}[section]
\newtheorem{theorem}{Theorem}[section]
\newtheorem{definition}{Definition}[section]
\newtheorem{proposition}{Proposition}[section]

%need to change
\def \be  {\begin{equation}}
\def \ee  {\end{equation}}
\def \al   {\alpha }
\def \la    {\lambda  }

\section{Introduction}

It is a usual practice to utilize the idea of linearization in analyzing 
%solution structures of 
nonlinear differential or differential-difference 
%soliton 
equations (see for example \cite{Calogero-book1990,BlumanK-EJAM1990}). 
The method of inverse scattering transform is an important application 
of such an idea to the theory of soliton equations \cite{GGKM-PRL1967,AblowitzS-book1981}, 
which has been recognized as one of the most significant contributions in the 
field of applied mathematics in the second half of the last century.    
The general formulation of 
Lax pairs is a spectacular tool of realization of inverse scattering transform
\cite{Lax-CPAM1968},
by which one can break a nonlinear problem into 
a couple of linear problems and then handle the resulting linear problems
to solve the nonlinear problem.
 
Recently in the past decade, an unusual way of using the nonlinearization technique arose in the theory of soliton equations 
\cite{Cao-SC1990}-\cite{Ma-book1990}. 
Although using the idea of nonlinearization is not normally 
considered to be  
a good direction in studying nonlinear equations,
one gradually realizes that the nonlinearization technique
provides a powerful approach for analyzing 
%nonlinear differential or differential-difference 
soliton equations, especially for showing the integrability by quadratures for soliton equations.
% because of the important role it plays \cite{Cao-SC1990,ZengL-JMP1990}.
%From a viewpoint of logical thinking, one can expect that 
%a negative thing of nonlinearity of soliton equations plus another negative thing of %nonlinearization   
%brings us a positive result.
The manipulation of nonlinearization
not only leads to finite-dimensional Liouville integrable systems 
\cite{Cao-SC1990}-\cite{Xu-CPL1995},
%,ZengL-JMP1990,Ma-book1990,AntonowiczW-JMP1992,RagniscoW-IP1992,Ma-AMAS1993,RagniscoW-JMP1994,
%Xu-CPL1995}, 
but also decomposes infinite-dimensional soliton equations, in whatever %$1+1$ and $2+1$ 
dimensions, into finite-dimensional Liouville integrable systems 
\cite{Ma-PA1995}-\cite{CaoWG-JMP1999}.
Moreover, it narrows the gap between infinite-dimensional soliton equations and 
finite-dimensional Liouville integrable systems 
\cite{AntonowiczW-JMP1992,Ma-PA1995,CaoWG-JMP1999},
and paves a method of separation of variables  
for soliton equations \cite{ZengM-JMP1999,MaZ-krustal2000},
which can also be used to analyze the resulting 
finite-dimensional integrable systems 
\cite{EilbeckEKT-JPA1994}-\cite{Blaszak-JMP1998}. 
Mathematically speaking, much excitement in the study of nonlinearization
comes from a kind of specific symmetry constraints 
\cite{KonopelchenkoSS-PLA1991}-\cite{Ma-JPSJ1995},
%,ChengL-PLA1991,MaS-PLA1994,Ma-JPSJ1995},
engendered from the variational derivative of the 
spectral parameter \cite{MaS-PLA1994,Ma-JPSJ1995}.
It is due to symmetry constraints that the nonlinearization technique
is so powerful in showing the integrability by quadratures 
for soliton equations \cite{MaFO-PA1996,MaDZL-NCB1996}.
The study of symmetry constraints itself is an important part of 
the kernel of the mathematical theory of nonlinearization,
which is also a common conceptional umbrella under which 
one can manipulate both mono-nonlinearization 
\cite{Cao-SC1990} and binary nonlinearization \cite{MaS-PLA1994}.

However,
%Nevertheless, 
all examples of application of the nonlinearization technique, discussed so far, 
are related to lower-order matrix (here, and in what follows, a
matrix is assumed to be square) spectral problems of soliton equations,
most of which are only concerned with 
second-order traceless matrix spectral problems.
On the one hand, there appears much difficulty  
in handling the Liouville integrability \cite{Arnold-book1978}
of the so-called constrained flows generated from spectral problems, 
in the case of the third-order and fourth-order matrix
spectral problems \cite{MaFO-PA1996,WuG-JMP1999,ShiZ-JMP2000}. 
It is a challenging task to extend the theory of nonlinearization
to the case of higher-order matrix spectral problems. 
On the other hand, one also notices that mono-nonlinearization can not be carried out in the 
cases of odd-order matrix spectral problems and even-order,
including the simplest second-order,
non-traceless matrix spectral problems.
Even for even-order traceless matrix spectral problems,
it is not clear how to determine pairs of canonical variables to obtain 
Hamiltonian structures of the constrained flows while doing mono-nonlinerization.
Therefore, one has to take into account adjoint spectral problems 
and manipulate binary nonlinearization for the case of
general matrix spectral problems. In the theory of binary
nonlinearization \cite{MaF-book1996}, there exists a natural way for determining symplectic
structures to exhibit Hamiltonian forms of the constrained flows.

In this paper, we 
would like to establish a concrete example to
apply the nonlinearization technique to the case of higher-order 
matrix spectral problems, by 
manipulating binary nonlinearization
for arbitrary-order matrix spectral problems
associated with the ${\cal N}$-wave interaction equations 
in both $1+1$ and $2+1$ dimensions. The resulting theory 
will show a direct way for generating sufficiently many integrals of motion, 
and more importantly
for proving the functional independence of the required integrals of motion, 
for the Liouville integrability of the constrained flows resulting from
higher-order matrix spectral problems. 

Let us recall some basic notation on binary nonlinearization (see, for example,
\cite{MaF-book1996} for a detailed description).
Let us assume that we have a matrix spectral problem  
\begin{equation}
 \phi_x=U\phi = U(u,\lambda)\phi,\ U=(U_{ij})_{r\times r},\ 
  \phi=(\phi_1,\cdots,\phi_r)^T
\label{gsp}
\end{equation}
with a spectral parameter $\la $ and a potential $u=(u_1,\cdots,u_q)^T$.
Suppose that the compatability conditions
\[ U_{t_m}-V_x^{(m)}+[U,V^{(m)}]=0,\ m\ge 0,\]
 of the spectral problem (\ref{gsp})
and the associated spectral problems
\be 
\phi_{t_m}=V^{(m)}\phi = V^{(m)}(u,u_x,\cdots;\lambda)\phi,\ V^{(m)}=(V^{(m)}_{ij})_{r\times r },\ m\ge 0, \label{gassosp}
\ee  
determine an isospectral ($\la _{t_m}=0$) soliton hierarchy
\be 
u_{t_m}=X_m(u)=JG_m=J\frac{\delta {\tilde H_m}}{\delta u},\ m\ge 0,
\label{sh}
\ee 
where $J$ is a Hamiltonian operator and ${\tilde H}_m$ are Hamiltonian functionals.
Obviously, the compatability conditions of 
the adjoint spectral problem
\be 
\psi _x=-U^T(u,\la )\psi ,\ \psi =(\psi _1,\cdots ,\psi_r)^T,
\ee 
and the adjoint associated spectral problems 
\be 
\psi _{t_m}=-V^{(m)T}\la =-V^{(m)T}(u,u_x,\cdots;\la )\psi  
\ee 
still give rise to the same hierarchy $u_{t_m}=X_m(u)$ defined by (\ref{sh}).
It has been pointed out \cite{MaS-PLA1994,Ma-PA1995} that 
$J\D\frac {\delta \lambda }{\delta u}$
is a common 
symmetry of all equations in the hierarchy (\ref{sh}).
Introducing 
$N$ distinct eigenvalues $\la _1,\la _2,\cdots,\la _N$,
we have 
\be 
\phi^{(s)}_x=U(u,\la _s)\phi^{(s)}, \ \psi^{(s)}_x=-U^T(u,\la _s)\psi^{(s)},
\ 1\le s\le N,
\label{gxpartofcf}
\ee 
and 
\be 
\phi^{(s)}_{t_m}=V^{(m)}(u,u_x,\cdots;\la _s)\phi^{(s)}, \ \psi^{(s)}_{t_m}
=-V^{(m)T}(u,u_x,\cdots;\la _s)\psi^{(s)},\ 1\le s\le N,
\label{gtpartofcf}
\ee 
where we set the corresponding eigenfunctions and adjoint eigenfunctions 
as $\phi^{(s)}$ and $\psi^{(s)}$, $1\le s\le N$.
It is assumed that the conserved covariant $G_{m_0}$ does not depend on 
any derivative of $u$ with respect to $x$, and thus 
the so-called general binary Bargmann symmetry constraint reads as   
\be
X_{m_0}=\sum_{s=1}^N E_s\mu _s J\frac {\delta \la _s}{\delta u},\ \textrm{i.e.,} 
\ JG_{m_0}=
J\sum_{s=1}^N\mu _s\psi^{(s)T}\frac {\part U(u,\la _s)}{\part u}\phi^{(s)},
\label{gsy} \ee 
where $\mu _s,\ 1\le s\le N,$ are arbitrary nonzero constants,
and $E_s,\ 1\le s\le N,$ are normalized constants.
The right-hand side of the symmetry constraint (\ref{gsy}) is 
a linear combination of $N$ symmetries 
\[
E_sJ\D\frac {\delta \lambda_s }{\delta u}=J\psi^{(s)T}\frac {\part U(u,\la _s)}{\part u}\phi^{(s)},\ 1\le s\le N.\]
Such symmetries are not Lie point, contact or Lie-B\"acklund
symmetries, since $\phi^{(s)}$ and $\psi^{(s)}$ can not be expressed 
in terms of $x$, $u$ and derivatives of $u$ with respect to $x$ to some finite order. 
%=J \D \psi^{(s)T}\frac {\part U(u,\la _s)}{\part u}\phi^{(s)},\ 1\le s\le N.
Suppose that (\ref{gsy}) has an inverse function
\be u=\widetilde u=\widetilde u(\phi^{(1)},\cdots,\phi^{(N)};\psi^{(1)},\cdots,\psi^{(N)}),
%((\delta F)  _1,\cdots,(\delta F) _q),\ (\delta F)_i=\sum_{s=1}^N\mu _s \psi^{(s)T}
%\frac {\part U(u,\la _s)}{\part u_i}\phi^{(s)},\ 1\le i\le q.
\ee 
Replacing $u$ with $\widetilde u$ in 
%$N$ replicas of   
the system ({\ref{gxpartofcf}) or the system (\ref{gtpartofcf}), 
we obtain the so-called spatial constrained flow:
\be 
\phi^{(s)}_x=U(\widetilde u,\la _s)\phi^{(s)}, \ \psi^{(s)}_x=-U^T(\widetilde u,\la _s)\psi^{(s)},
\ 1\le s\le N,
\label{xpartofcf}
\ee 
or the so-called temporal constrained flows:
\be 
\phi^{(s)}_{t_m}=V^{(m)}(\widetilde u,\widetilde u_x,\cdots;\la _s)\phi^{(s)}, \ \psi^{(s)}_{t_m}
=-V^{(m)T}(\widetilde u,\widetilde u_x,\cdots;\la _s)\psi^{(s)}, \ 1\le s\le N.
\label{tpartofcf}
\ee 
The main problem of nonlinearization is to show that 
the spatial constrained flow (\ref{xpartofcf})
and the temporal constrained flows (\ref{tpartofcf}) 
under the control of (\ref{xpartofcf})
are Liouville integrable. 
Then if $\phi^{(s)}$ and $\psi ^{(s)}$, $1\le s\le N$,
solve two constrained flows (\ref{xpartofcf}) and (\ref{tpartofcf}) simultaneously,  
 $u=\widetilde u$
will give rise to a solution to 
the $m$th soliton equation $u_{t_m}=X_m(u)$. It also follows that the soliton equation 
$u_{t_m}=X_m(u)$ is decomposed into two finite-dimensional Liouville integrable 
systems, and $u=\widetilde u$ presents a B\"acklund transformation 
between infinite-dimensional soliton equations and 
finite-dimensional Liouville integrable systems.
More generally, if a soliton equation is associated with 
a set of spectral problems 
\[ \phi_{x_i}=U^{(i)}(u,\la )\phi,\ 1\le i\le p,\]
then it will be decomposed into $p+1$ 
finite-dimensional Liouville integrable systems. 
The above whole process is called binary nonlinearization \cite{Ma-PA1995,MaF-book1996}.
%For binary nonlinearization of discrete soliton equations, please refer to \cite{MaG-book1999}.

This paper is structured as follows.
In Section 2, we will present 
binary symmetry constraints of the ${\cal N}$-wave interaction equations
in $1+1 $ dimensions,
and show Hamiltonian structures and Lax presentations of the corresponding constrained flows.
In Section 3, we consider the $2+1$ dimensional case.
We will similarly construct 
binary symmetry constraints of the ${\cal N}$-wave interaction equations
in $2+1 $ dimensions, and discuss some properties of the corresponding constrained flows. 
In Section 4, we go on to propose an involutive system 
of functionally independent polynomial functions,
generated from an arbitrary-order matrix Lax operator,
along with an alternative involutive and functionally independent system. 
An ${\bf r}$-matrix formulation will be established for the Lax operator, 
and used to show the involutivity of the obtained system of polynomial functions,
together with Newton's identities on elementary symmetric polynomials.
A detailed proof will also be made 
for the functional independence of the system of polynomial functions
by using the determinant property of the tensor product of matrices.
In Section 5, two applications of the involutive system engendered in Section 4
will be given, which verify that all constrained flows 
associated with the ${\cal N}$-wave interaction equations 
in both $1+1$ and $2+1$ dimensions are Liouville integrable.
Moreover, a kind of 
involutive solutions of the ${\cal N}$-wave interaction equations in two cases
will be depicted.
Theses also show the integrability by quadratures for the ${\cal N}$-wave interaction 
equations. Finally in Section 6, some concluding remarks will be given, together with 
conclusions.

\section{Binary symmetry  constraints %of the ${\cal N}$-wave interaction equations
in $1+1 $ dimensions}\label{subsec:constraint11}
\setcounter{equation}{0}

\subsection{$n\times n$ AKNS hierarchy and
$1+1$ dimensional ${\cal N}$-wave interaction equations}  

Let $n$ be an arbitrary natural number strictly greater than two.
We begin with the $n\times n$ matrix AKNS spectral problem 
\cite{NovikovMPZ-book1984}  
\begin{equation}
 \phi_x=U\phi = U(u,\lambda)\phi,\
 U(u,\lambda)=\lambda U_0+U_1(u),\  \phi=(\phi_1,\cdots,\phi_n)^T,
\label{spofnWIEsin1+1}
\end{equation}
with a spectral parameter $\la $ 
and 
\be U_0=\textrm{diag}(\alpha_1,\cdots,\alpha_n),\ U_1(u)=
(u_{ij})_{n\times n},\ee
where $\al _i$, $1\le i\le n$, are distinct constants,
and $u_{ii}=0,\ 1\le i\le n.$ 
The standard AKNS spectral problem, 
i.e., the spectral problem (\ref{spofnWIEsin1+1}) with $n=2$,
has been analyzed in \cite{AblowitzKNS-SAM1974}, but it
can not generate any ${\cal N}$-wave interaction equations and thus it is not discussed
here. 
In order to express related soliton equations in a compact form, 
we write down the potential $u$ as 
\be \left\{\ba {l}
 u=\rho (U),\  \textrm{i.e.,} \ 
%u=(u_{21},u_{12})^T,\ \textrm{when}\ n=2,\vspace{2mm}\\
u=(u_{21},u_{12},u_{13},u_{31},
u_{23},u_{32})^T,\ \textrm{when}\ n=3,\vspace{2mm}\\
u=(u_{21},u_{12},u_{13},u_{31},
u_{14},u_{41},u_{23},u_{32},\cdots, u_{n,n-1},u_{n-1,n})^T,\ \textrm{when}\ n\ge 4,
\ea \right.
\label{eq:rho}
\ee 
in which we arrange the exponents $u_{ij}$ in a specific way, 
first from smaller to larger 
of the integers $k=i+j$ and then symmetrically for each set 
$\{u_{i,k-i}|1\le i\le k-1\}$. 

Let us now consider the construction of the $1+1$ dimensional ${\cal N}$-wave interaction equations
and its whole isospectral hierarchy  
associated with the spectral problem (\ref{spofnWIEsin1+1}). 
We first solve the stationary zero-curvature equation for $W$: 
\be W_x-[U,W]=0 ,\  W=(W_{ij})_{n\times n},\label{zcein1+1} \ee 
which is equivalent to 
\be \left\{
\ba {l} W_{ij,x}+u_{ij}(W_{ii}-W_{jj})+
{\displaystyle 
\sum^n_{\stackrel{k=1}{k\not=i,j}}}
(u_{kj}W_{ik}-u_{ik}W_{kj})-\lambda(\alpha_i-\alpha_j)W_{ij}=0,\ i\not=j, \vspace{2mm} \\
W_{ii,x}=
{\displaystyle 
\sum^n_{\stackrel{k=1}{k\not=i}}}
(u_{ik}W_{ki}-u_{ki}W_{ik}),
\ea \right. \label{eeofzcein1+1}
\ee 
where $1\le i,j\le n$.
We look for a formal solution of the form
\be W=\sum_{l\geq 0}W_{l}\lambda^{-l},\ 
W_{l}=(W^{(l)}_{ij})_{n\times n} ,\label{formofW_{ij}}\ee 
and thus (\ref{eeofzcein1+1}) becomes the following recursion relation
\be 
\left \{\ba {l} 
W^{(0)}_{ii,x}=0,\  W^{(0)}_{ij}=0,\  i\not=j,\vspace{2mm}\\
W^{(l)}_{ij,x}+u_{ij}(W^{(l)}_{ii}-W^{(l)}_{jj})+
{\displaystyle 
\sum^n_{\stackrel{k=1}{k\not=i,j}}}
(u_{kj}W^{(l)}_{ik}-u_{ik}W^{(l)}_{kj})-(\alpha_i-\alpha_j)W^{(l+1)}_{ij}=0,\  i\not=j,
\vspace{2mm}\\
W^{(l+1)}_{ii,x}=
{\displaystyle 
\sum^n_{\stackrel{k=1}{k\not=i}}}
(u_{ik}W^{(l+1)}_{ki}-u_{ki}W^{(l+1)}_{ik}),\ea \right. \label{rrofW_{ij}}
\ee 
where $1\le i,j\le n$ and $l\geq 0$.
In particular, from the above recursion relation, we have that 
\be W^{(0)}_{ii}=\beta _i=\textrm{const.}, \  W^{(0)}_{ij}=0,\ 1\le i\not=j\le n,
\ee 
and 
\be 
W^{(1)}_{ii}=0,\ W^{(1)}_{ij}=\frac{\beta_i-\beta_j}{\alpha_i-\alpha_j}u_{ij},\  1\le 
i\not=j\le n.
\ee 
We require that 
\be 
 W^{(l)}_{ij}|_{u=0}=0,\  1\le i,j\le n,\ l\geq 1.
\label{requirementofuniques}
\ee 
This condition (\ref{requirementofuniques}) means to identify all constants of integration 
to be zero while using (\ref{rrofW_{ij}}) to determine $W$, and thus 
all $W_{l},\ l\ge 1, $ will be uniquely determined. 
For example, we can obtain from (\ref{rrofW_{ij}}) under 
(\ref{requirementofuniques}) that
\be \left\{\ba {l}
W^{(2)}_{ij}={\displaystyle \frac{\beta_i-\beta_j}{(\alpha_i-\alpha_j)^2}}
u_{ij,x}+{\displaystyle \frac{1}{\alpha_i-\alpha_j}}
{\displaystyle 
\sum^n_{\stackrel{k=1}{k\not=i,j}}
(\frac{\beta_k-\beta_i}{\alpha_k-\alpha_i}-\frac{\beta_k-\beta_j}{\alpha_k-\alpha_j})u_{ik}u_{kj},\  1\le i\not=j\le n,}\vspace{2mm}\\
W^{(2)}_{ii}=
{\displaystyle 
\sum^n_{\stackrel{k=1}{ k\not=i}}
\frac{\beta_k-\beta_i}{(\alpha_k-\alpha_i)^2}u_{ik}u_{ki}}, \ 1\le i\le n. \ea \right.
\label{W^{(2)}} \ee  
It is easy to see that the recursion relation (\ref{rrofW_{ij}}) can lead to
\be \ba{l}
2u_{ij}\partial^{-1}u_{ij}W^{(l)}_{ji}+(\partial-2u_{ij}\partial^{-1}u_{ji})W^{(l)}_{ij}+
{\displaystyle 
\sum^n_{\stackrel{k=1}{k\not=i,j}}}
\left[u_{ij}\partial^{-1}u_{ik}W^{(l)}_{ki}+(u_{kj}-u_{ij}\partial^{-1}u_{ki})W^{(l)}_{ik}\right]
\vspace{2mm}\\
\quad {\displaystyle 
+\sum^n_{\stackrel{k=1}{k\not=i,j}}}
\left[u_{ij}\partial^{-1}u_{kj}W^{(l)}_{jk}-(u_{ik}+u_{ij}\partial^{-1}u_{jk})W^{(l)}_{kj}\right]=(\alpha_i-\alpha_j)W^{(l+1)}_{ij},\ 
 i\not=j, \ea 
%\label{anotherformofrrofW_{ij}}
\ee 
where $1\le i,j\le n$, $ l\ge 1$, and $\partial ^{-1}$ is the inverse operator of $\partial =\frac {\partial }{\partial x}$.
This can be written as the Lenard form
\be MG_{l-1}=JG_l,\  l\geq 1, \label{rrofG_l}\ee 
where $G_{l}=\rho (W_{l+1})$
is generated from $W_{l+1}$ in the same way as that for $u$, and 
$J$ is a constant operator 
\be \left\{\ba {l}
%J=(\al _1-\al _2)\sigma_0,\ \textrm{when}\ n=2,\vspace{2mm}\\
J=\textrm{diag}\Bigl( (\alpha_1-\alpha_2)\sigma_0, (\alpha_1-\alpha_3)\sigma_0,(\alpha_2-\alpha_3)\sigma_0\Bigl),
\ \textrm{when}\ n=3,\vspace{2mm}\\
 J=\textrm{diag}
\Bigl( \underbrace{(\alpha_1-\alpha_2)\sigma_0, (\alpha_1-\alpha_3)\sigma_0,
(\alpha_1-\alpha_4)\sigma_0,
(\alpha_2-\alpha_3)\sigma_0,\cdots,(\alpha_{n-1}-\alpha_n)\sigma_0}
_{n(n-1)/2}\Bigr),
\vspace{2mm}\\
\qquad \qquad \qquad \qquad \qquad \qquad \qquad \qquad \qquad 
\qquad \qquad \qquad \qquad \qquad \qquad \textrm{when}\ n\ge 4,
 \ea \right.
\label{defofJ}
\ee 
with $\sigma_0$ being given by 
\[ \sigma_0=\left( \begin{array} {cc} 0&1 \vspace{2mm} \\ -1&0 \end{array}
 \right ). \]  
For example,
when $n\ge 4$, we have
\be 
G_{l-1}=
(W^{(l )}_{21},W^{(l )}_{12},W^{(l )}_{31},W^{(l )}_{13},W^{(l )}_{41},W^{(l )}_{14},W^{(l)}_{32},
W^{(l )}_{23},\cdots,W^{(l)}_{n,n-1},W^{(l )}_{n-1,n})^T,
\ l\ge 1,\ee 
the first of which reads as 
\bea  
G_0&=&\Bigl(\Bigr.\frac{\beta_1-\beta_2}{\alpha_1-\alpha_2}u_{21},\frac{\beta_1-\beta_2}{\alpha_1-\alpha_2}u_{12},\frac{\beta_1-\beta_3}{\alpha_1-\alpha_3}u_{31},
\frac{\beta_1-\beta_3}{\alpha_1-\alpha_3}u_{13},
\frac{\beta_1-\beta_4}{\alpha_1-\alpha_4}u_{41},
\frac{\beta_1-\beta_4}{\alpha_1-\alpha_4}u_{14},
\nonumber \\
&& \cdots,\frac{\beta_{n-1}-\beta_n}{\alpha_{n-1}-\alpha_n}u_{n,n-1},\frac{\beta_{n-1}-\beta_n}{\alpha_{n-1}-\alpha_n}u_{n-1,n}\Bigl.\Bigr)^T. 
\label{defofG_0}
\eea  
The operators $J$ and $M$ are skew-symmetric and can be shown to be a Hamiltonian pair
\cite{GelfandD-FAP1979,FuchssteinerF-PD1981}.

We proceed to introduce the associated spectral problems with the spectral problem (\ref{spofnWIEsin1+1})
\be 
\phi_{t_m}=V^{(m)}\phi,\  V^{(m)}=V^{(m)}(u,\lambda)=(\lambda^mW)_+,\  m\geq 1,
\label{aspofmthnWIEsin1+1}
\ee 
where the symbol $+$ stands for the choice of the part of non-negative powers of $\lambda$. 
Note that we have 
\[ W_{lx}=[U_0,W_{l+1}]+[U_1,W_l],\ l\ge0, \]
and we can compute that 
\bea  && [U,V^{(m)}]=[\lambda U_0+U_1,\sum_{l =0}^m \lambda ^{m-l }W_l ]\nonumber \\
&=& \sum_{l =0}^m [U_0,W_l ]\lambda ^{m+1-l }+\sum_{l =0}^m[U_1,W_l ]\lambda ^{m-l }\nonumber \\
&=& \sum_{l =0}^{m-1}[U_0,W_{l +1}]\lambda ^{m-l }
+\sum_{l =0}^m[U_1,W_l ]\lambda ^{m-l },
\nonumber \eea 
where we have used $[U_0,W_0]=0$.
Therefore, under the isospectral conditions 
\be \la _{t_m}=0,\ m\ge 1,\ee 
the compatibility conditions of the spectral problem 
(\ref{spofnWIEsin1+1}) and the associated spectral problems 
(\ref{aspofmthnWIEsin1+1}), i.e.,
the zero-curvature equations 
\[U_{t_m}-V^{(m)}_x+[U,V^{(m)}]=0,\ m\ge 1,\]
equivalently lead to  
\[ U_{1t_m}=W_{mx}-[U_1,W_m]=[U_0,W_{m+1}],\ m\ge 1.
\] 
This gives rise to the so-called $n\times n$ AKNS soliton hierarchy  
\be 
u_{t_m}=X_m :=JG_m , \ m\geq 1,
\label{mthnWIEsin1+1}
%\eqno(2.13)
\ee 
where $J$ and $G_m=\rho (W_{m+1})$ are determined by (\ref{defofJ}) 
and (\ref{rrofG_l}).

Applying the trace identity  \cite{Tu-JMP1989} 
\[ \frac \delta {\delta u }\int \textrm{tr}(W\frac {\part U}{\part \la })dx=
\la ^{-\gamma }\frac {\part }{\part \la }\la ^\gamma \textrm{tr}(W\frac {\part U}{\part u})
 \]
where $\gamma $ is a constant to be determined,
we can obtain
\be 
\frac{\delta \tilde H_l}{\delta u_{ij}}=W^{(l)}_{ji},\  \tilde H_l:=-\frac{1}{l}\int
(\alpha_1W^{(l+1)}_{11}+\alpha_2W^{(l+1)}_{22}+\cdots+\alpha_nW^{(l+1)}_{nn})dx,\ l\ge 1,
\ee 
in which $1\le i\ne j\le n$ and $\gamma $ is determined to be zero.
In this computation, we need to note that
\[ \textrm{tr}(W\frac {\part U}{\part \la }) =
\textrm{tr}(WU_0)=\sum_{l\ge 0}(\al _1W_{11}^{(l)}+\al _2W_{22}^{(l)}+\cdots +
\al _nW_{nn}^{(l)})\la ^{-l}, 
\]
and 
\[ \textrm{tr}(W\frac {\part U}{\part u_{ij}})
=\textrm{tr}(W E_{ij})=W_{ji}=\sum_{l\ge 0}W_{ji}^{(l)}\la ^{-l},  \ 1\le i\ne j\le n,
\]
where $E_{ij}$ is an $n\times n$ matrix whose $(i,j)$ entry 
is one but other entries are all zero.
Therefore, 
the isospectral hierarchy (\ref{mthnWIEsin1+1})
has a bi-Hamiltonian formulation
\be u_{t_m}=X_m =J \frac{\delta \tilde H_{m+1}}{\delta u}=M \frac{\delta \tilde 
H_{m}}{\delta u}, \ m\geq 1. \ee 

%For fixed $m$, each set of $\{c_i\}$ in (\ref{eq:def_ci}) gives a
%system of nonlinear partial differential equations of $u$.

%Let $\cc_i=\beta_i$ $(i=1,\cdots,n)$, the
The first nonlinear system in the hierarchy 
(\ref{mthnWIEsin1+1}) is the $1+1$ dimensional ${\cal N}$-wave interaction equations 
\cite{AblowitzH-JMP1975}
\begin{equation}
u_{ij,t_1}=\frac{\beta_i-\beta_j}{\alpha_i-\alpha_j}u_{ij,x}
+\sum^n_{\stackrel{k=1}{k\not=i,j}}(\frac{\beta_i-\beta_k}{\alpha_i-\alpha_k}
-\frac{\beta_k-\beta_j}{\alpha_k-\alpha_j})u_{ik}u_{kj},\  1\leq i\ne j\leq n.
\label{nWIEsin1+1}
\end{equation}
This system is actually equivalent to the following equation in the matrix form
\be U_{1t_1}=W_{1x}-[U_1,W_1],  \label{mfofNWIEs1+1}\ee
%where $U_1$ and $W_1$ are defined as before.
which can be rewritten as 
\be P_{t_1}=Q_x-[P,Q],\ [U_0,Q]=[W_0,P], \ee  
where $P$ and $Q$ are assumed to be two off-diagonal potential matrices.
Based on (\ref{mfofNWIEs1+1}), a vector field $\rho (\delta P)$ is a symmetry of (\ref{nWIEsin1+1}) if the matrix $\delta P $ satisfies 
the linearized system of (\ref{nWIEsin1+1}):
\be 
(\delta P) _{t_1}=(\delta Q) _x-[U_1,\delta Q ]  -[\delta P ,W_1]
\label{lsofNWIEs1+1}
\ee  
with $\delta Q$ being determined by 
\be [U_0,\delta Q ]=[W_0,\delta P ]. 
\label{alsofNWIEs1+1}
\ee 

The ${\cal N}$-wave interaction equations (\ref{nWIEsin1+1})
contains a couple of 
physically important nonlinear models as special reductions \cite{AlbowitzC-book1991},
for example, three-wave interaction equations
arising in fluid dynamics and plasma physics 
\cite{ZakhrovM-SPJEPTL1973,Kaup-SAM1976,EnnsM-book1997}, 
with $U$ being chosen to be an anti-Hermitian matrix. 
Its Darboux transformation has been 
established in \cite{Gu-book1990},
which allows one to construct 
soliton solutions in a purely algebraic way.
The Darboux transformation has also been analyzed 
for the ${\cal N}$-wave interaction equations with additional linear terms
\cite{Leble-CMA1998}.

\subsection{Binary symmetry constraints in $1+1$ dimensional case}  

We would like to present binary symmetry constraints of the $1+1$ dimensional 
${\cal N}$-wave interaction equations (\ref{nWIEsin1+1}).
To this end, we need to introduce
the adjoint spectral problem of 
(\ref{spofnWIEsin1+1}):
\be 
\psi_x=-U^T(u,\lambda)\psi, \ \psi=(\psi_1, \cdots, \psi_n)^T,
\label{aspofnWIEsin1+1}
%\eqno(3.1)
\ee 
and the adjoint 
associated spectral problem of (\ref{aspofmthnWIEsin1+1}):
\be 
\psi_{t_m}=-V^{(m)T}(u,\lambda)\psi, 
\label{aaspofmthnWIEsin1+1}
\ee 
where $U$ and $V^{(m)}$ are given as in (\ref{spofnWIEsin1+1}) and 
(\ref{aspofmthnWIEsin1+1}), respectively.
The compatability condition of (\ref{aspofnWIEsin1+1}) and (\ref{aaspofmthnWIEsin1+1})
still gives rise to  
$u_{t_m}=X_m$ defined by (\ref{mthnWIEsin1+1}).
%Therefore, the ${\cal N}$-wave interaction equation (\ref{nWIEsin1+1})\
%has the following zero equation representation
%\be  \ee  
 
The variational derivative of the spectral parameter $\lambda $ with respect to the potential
$u$ can be calculated by (see \cite{MaS-PLA1994,MaFO-PA1996}, or
\cite{Ma-PA1995} for a detailed deduction)
\be \frac {\delta \lambda }{\delta u}= E^{-1}\psi ^T\frac {\part U}{\part u}\phi,
\ \textrm{i.e.,}\ 
\frac {\delta \lambda }{\delta u_{ij}}=E^{-1}\phi _i\psi_j,\ 1\le i\ne j\le n,
\ee
where $E$ is the normalized constant:
\[ E=-\int _{-\infty}^\infty \psi ^T\frac {\part U}{\part \lambda }\phi dx.\]
A direct calculation can show that the 
variational derivative 
 satisfies the following equation
\be 
M \frac{\delta  \lambda }{\delta u}=\lambda  J \frac {\delta \lambda}
{\delta u} . %\eqno(3.4)
\ee 
Since $\lambda $ does not vary with respect to time,
we have a specific common symmetry 
$J\frac{\delta \la }{\delta u}$ of the hierarchy (\ref{mthnWIEsin1+1}).
To carry out binary nonlinearization, 
we take a Lie point symmetry of 
the $\cal N$-wave interaction equations
(\ref{nWIEsin1+1}), 
\be 
Y_0:=\rho ( [\Gamma, U_1 ]  ),\ \Gamma =\diag ( \gamma _1,\cdots , \gamma_n),
\ee  
where $\gamma_1,\gamma_2,\cdots ,\gamma_n$ are arbitrary distinct constants ($X_0=JG_0$ is an example
with $\Gamma =W_0$).
It can be easily checked that 
\[ (\delta P,\delta Q)=([\Gamma , U_1],[\Gamma , W_1])\]
satisfies (\ref{lsofNWIEs1+1}), and thus $Y_0$ is a symmetry of (\ref{nWIEsin1+1}).
Then, make the following binary Bargmann symmetry constraint 
\be Y_0=\mu E  J\frac {\delta \lambda }{\delta u}=\mu  J
\psi ^T\frac {\part U}{\part u}\phi,
%\label{originalgBscofnWIEsin1+1}
\ee 
where $\mu $ is an arbitrary nonzero constant,
$J$ is defined by (\ref{defofJ}), and
$\phi$ and $\psi$ are the eigenfunction and adjoint eigenfunction 
of (\ref{spofnWIEsin1+1}) and (\ref{aspofnWIEsin1+1}), respectively.
Upon introducing $N$ distinct eigenvalues
 $\lambda_1,\lambda_2,\cdots,\lambda_N$,  
we obtain a general binary symmetry constraint
\be 
Y_0= J\sum _{s=0}^N \mu _s 
\psi ^{(s)T}\frac {\part U(u,\lambda _s)}{\part u}\phi ^{(s)}:=Z_0,
\label{eq:originalgBscofnWIEsin1+1}
%due to revision \label{gBscofnWIEsin1+1}
\ee 
where $\mu _s$, $1\le s\le N$, are $N$ nonzero constants, and
$\phi^{(s)}$ and $\psi^{(s)}$, $ 1\le s\le N$, are eigenfunctions and 
adjoint eigenfunctions defined by 
\be 
\phi ^{(s)}_x=U(u,\lambda _s)\phi ^{(s)},\ 
\psi ^{(s)}_x=-U^T(u,\lambda _s)\psi ^{(s)},\ 1\le s\le N,
\label{Nreplicasofspandasp}
\ee 
and
\be
\phi ^{(s)}_{t_1}=V^{(1)}(u,\la _s)\phi^{(s)},\ \psi^{(s)}_{t_1}=-V^{(1)T}(u,\la _s)\psi^{(s)},
\ 1\le s\le N.\label{NreplicasofspofV^{(1)}}
\ee  

Let us rewrite the left-hand side of (\ref{eq:originalgBscofnWIEsin1+1})
as the matrix form
\be \delta P=\rho^{-1}(Z_0)=[U_0,\sum_{s=1}^N\mu_s \phi^{(s)}\psi^{(s)T}], \ee
which allow us to prove, by a direct computation as in \cite{MaL-PLA2000}
but more conveniently, that the vector field $Z_0=\rho(\delta P)$ 
%corresponding to this matrix $\delta P$ 
is really a symmetry of 
the $\cal N$-wave interaction equations (\ref{nWIEsin1+1}).
Now the symmetry problem is equivalent to showing that 
\be 
(\delta P ,\delta Q )=([U_0,\sum_{s=1}^N\mu_s \phi^{(s)}\psi^{(s)T}],
[W_0,\sum_{s=1}^N\mu_s \phi^{(s)}\psi^{(s)T}])
%\label{2ndsymmetryofNWIEs1+1}
\ee 
satisfies the linearized system (\ref{lsofNWIEs1+1}),
when $\phi^{(s)}$ and $\psi^{(s)}$, $1\le s\le N$, satisfy 
(\ref{Nreplicasofspandasp}) and (\ref{NreplicasofspofV^{(1)}}).
A detailed proof will be given in Appendix A.

Therefore, we have the following binary symmetry constraint
\be 
Y_0= J\sum _{s=0}^N \mu _s 
\psi ^{(s)T}\frac {\part U(u,\lambda _s)}{\part u}\phi ^{(s)},\ \textrm{i.e.}, \ 
[\Gamma, U_1]=[U_0, \sum_{s=1}^N\mu_s \phi^{(s)}\psi^{(s)T}].
\label{gBscofnWIEsin1+1}
\ee 
When $N$ and $\mu _s $ vary, (\ref{gBscofnWIEsin1+1})
provides us with a set of binary symmetry constraints of 
the ${\cal N}$-wave interaction equations (\ref{nWIEsin1+1}). 
%From two sets of constants $\{\la _s\}_{s=1}^N$ and $\{\mu _s\}_{s=1}^N$,
%we form two diagonal matrices
%\be A=\textrm{diag}(\la _1,\la _2,\cdots,\la _N),\ B=
%\textrm{diag}(\mu _1,\cdots,\mu _N),\label{defofAB}\ee 
%which will be used throughout our discussion.
Let us assume that 
%\be \beta _i\ne \beta _j,\ 1\le i\ne j\le n,\ee
%and 
\be 
\phi ^{(s)}= (\phi _{1s},\phi_{2s},\cdots, \phi_{ns})^T,
\ \psi ^{(s)}= (\psi _{1s},\psi_{2s},\cdots, \psi_{ns})^T,
\ee 
in order to get an explicit expression for $u$ from
the symmetry constraint (\ref{gBscofnWIEsin1+1}), and 
introduce two diagonal matrices
\be A=\textrm{diag}(\la _1,\cdots,\la _N),\ B=
\textrm{diag}(\mu _1,\cdots,\mu _N),\label{defofAB}
\ee 
which will be used throughout our discussion. 
Solving the Bargmann symmetry constraint (\ref{gBscofnWIEsin1+1}) for $u$, 
we obtain
\be 
u_{ij}=\widetilde u_{ij}:=\frac{\alpha_i-\alpha_j}{\gamma_i-\gamma_j}\langle\Phi_i,B\Psi_j\rangle,\ 
  1\leq i\ne j\leq n, 
\label{componentofgBscofnWIEsin1+1}
\ee 
where $B$ is given by (\ref{defofAB}), and $\Phi _i$ and $\Psi _i$ are defined by
\be \Phi_i=(\phi_{i1},\phi_{i2},\cdots,\phi_{iN})^T,\ 
\Psi_i=(\psi_{i1},\psi_{i2},\cdots,\psi_{iN})^T,\ 
1\leq i\leq n, \label{defofPhi_iPsi_i}
\ee 
and $\langle\cdot,\cdot\rangle$ denotes the standard inner-product of the Euclidean space ${\R}^N$. 

Note that the compatability condition of (\ref{Nreplicasofspandasp}) and 
(\ref{NreplicasofspofV^{(1)}})
is still nothing but the $1+1$ dimensional ${\cal N}$-wave interaction equations 
(\ref{nWIEsin1+1}).
Now using (\ref{componentofgBscofnWIEsin1+1}),
we nonlinearize the spatial part (\ref{Nreplicasofspandasp}) and the temporal part 
(\ref{NreplicasofspofV^{(1)}}) of 
spectral problems and adjoint spectral problems of 
the ${\cal N}$-wave interaction equations (\ref{nWIEsin1+1}).
Namely we replace $u_{ij}$ with $\widetilde u_{ij}$
in $N$ replicas of the spectral problems 
and adjoint spectral problems (\ref{Nreplicasofspandasp})
and $N$ replicas of the associated spectral problems 
and adjoint associated spectral problems (\ref{NreplicasofspofV^{(1)}}),
and then obtain two constrained flows for the ${\cal N}$-wave interaction equations 
(\ref{nWIEsin1+1}):
\be 
\phi ^{(s)}_x=U(\widetilde u, \lambda _s)\phi ^{(s)}, \  \psi ^{(s)}_x=-U^T(\widetilde u, \lambda _s)\psi ^{(s)},\ 1\le s\le N,
\label{spatialconstrainedflowin1+1}
\ee 
and 
\be 
\phi ^{(s)}_{t_1}=V^{(1)}(\widetilde u, \lambda _s)\phi ^{(s)}, \  \psi ^{(s)}_{t_1}
=-V^{(1)T}(\widetilde u, \lambda _s)\psi ^{(s)},\ 1\le s\le N,
\label{temporalconstrainedflowin1+1}
\ee 
where $\widetilde u=\rho ((\widetilde u_{ij})_{n\times n})$ is defined like $u$.
For example, when $n\ge 4$, we have
\be \widetilde u=(\widetilde u_{21},\widetilde u_{12},\widetilde u_{31},\widetilde u_{13},\widetilde u_{14},
\widetilde u_{41},\widetilde u_{23},\widetilde u_{32},\cdots, \widetilde u_{n,n-1},\widetilde u_{n-1,n})^T.\ee 

In order to analyze the Liouville integrability of the above two constrained flows,
let us first introduce
a symplectic structure
\be 
\omega ^2= \sum_{i=1}^n B d\Phi _i\wedge d\Psi _i=
\sum_{i=1}^n\sum_{s=1}^N \mu _s  d\phi_{is}\wedge d\psi _{is}
\label{symplecticform}
\ee 
over $\R ^{2nN}$,
and then the corresponding Poisson bracket 
\bea && \{f,g\}=\omega ^2 (Idg,Idf)=\sum_{i=1}^n(\langle \frac {\part f}{\part \Psi_i},
B^{-1}\frac {\part g}{\part \Phi_i} \rangle -
\langle \frac {\part f}{\part \Phi_i},
B^{-1}\frac {\part g}{\part \Psi_i} \rangle  )\nonumber \\ &&
=\sum_{i=1}^n\sum_{s=1}^N
\mu _s^{-1} \Bigl(\frac{\part f}{\part \psi_{is}} \frac{\part g}{\part \phi_{is}}
-\frac{\part f}{\part \phi_{is}} \frac{\part g}{\part \psi_{is}}
\Bigr),\ f,g\in C^\infty (\R ^{2nN}), \label{eq:Poissonbracketin1+1case}\eea  
where the vector field $Idf$ is defined by 
\[\omega ^2(X,Idf)=df(X),\ X\in T(\R ^{2nN}).  \]
A Hamiltonian system with a Hamiltonian $H$ defined over 
the symplectic manifold $(\R ^{2nN},\omega ^2)$ is given by 
\be 
\Phi_{it}=\{\Phi_{i},H\}=-B^{-1}\frac {\part H}{\part \Psi _i},\ \Psi_{it}=
\{\Psi_{i},H\}=
B^{-1}\frac {\part H}{\part \Phi_i},\ 1\le i\le n,
\label{eq:gHamiltonianform}
\ee 
where $t$ is assumed to be the evolution variable.
Second, we need a matrix Lax operator 
\be 
L^{(1)}(\lambda )=C_1 + D_1(\la ),\label{defofL^{(1)}}\ee 
with $C_1$ and $D_1(\la )$ being defined by
\be  
C_1=\Gamma=\textrm{diag}(\gamma _1,\cdots,\gamma_n),\  
D_1(\la )=(D^{(1)}_{ij}(\la ))_{n\times n},\ D^{(1)}_{ij}(\la )=
\sum_{s=1}^N \frac {\mu _s} {\lambda -\la _s} \phi_{is}\psi_{js},\label{defofL^{(1)}c}\ee 
where $1\le i,j\le n$. 
Note that upon taking binary nonlinearization,
we obtain
\bea && U(\widetilde u,\lambda )= \lambda U_0+U_1(\widetilde u)
   =\lambda U_0+(\widetilde u_{ij}),\ \widetilde u_{ij}
   =\frac {\al _i-\al _j}{\gamma _i-\gamma _j}
\langle\Phi_i,B\Psi_j\rangle,\label{defofU(f,lambda)}
 \\ &&
V^{(1)}(\widetilde u,\lambda )= \lambda W_0 +W_1(\widetilde u)=\lambda W_0+( \widetilde v_{ij})
,\ \widetilde v_{ij}:=\frac {\beta _i-\beta  _j}{\al  _i-\al  _j}\widetilde u_{ij}
=\frac {\beta _i-\beta  _j}{\gamma _i-\gamma  _j}\langle\Phi_i,B\Psi_j\rangle,\qquad \label{defofV^{(1)}(f,lambda)} \quad 
\label{eq:V(1)}
\eea  
where $1\le i,j\le n$.

\begin{theorem}\label{thm:Hamiltoniansin1+1}
Under the symplectic structure (\ref{symplecticform}),
the spatial constrained flow (\ref{spatialconstrainedflowin1+1})
and the temporal constrained flow (\ref{temporalconstrainedflowin1+1})
for the $1+1$ dimensional ${\cal N}$-wave interaction equations 
(\ref{nWIEsin1+1})
are Hamiltonian systems with the evolution variables $x$ and ${t_1}$, and the Hamiltonians 
\bea && H^x_1= -\sum^n_{k=1}\alpha_k\langle A    \Phi_k,B\Psi_k\rangle-
\sum_{1\leq k<l\leq n}\frac{\alpha_k-\alpha_l}{\gamma_k-\gamma_l}\langle\Phi_k,B\Psi_l\rangle\langle\Phi_l,B\Psi_k\rangle,\label{H^x_1}\qquad \\
&& H^{t_1}_1= -\sum_{k=1}^n\beta _k\langle A   \Phi _k,B\Psi_k\rangle-\sum_{1\le k<l\le n}
\frac{\beta _k-\beta _l}{\gamma_k-\gamma_l}
\langle\Phi _k,B\Psi _l\rangle\langle\Phi _l,B\Psi_k\rangle,\label{H^t_1}
\eea 
respectively,
where $A$ and $B$ are defined by (\ref{defofAB}), and 
$\Phi _i$ and $\Psi_i$, $1\le i\le n$, are defined by (\ref{defofPhi_iPsi_i}).
%$A  =\textrm{diag}(\lambda _1,\cdots, \lambda _N),$
Moreover, they possess necessary Lax representations, i.e., we have
\be 
(L^{(1)}(\lambda ))_x=[U(\widetilde u,\lambda ), L^{(1)}(\lambda )],
\ (L^{(1)}(\lambda ))_{t_1}=  [V^{(1)}(\widetilde u,\lambda ), L^{(1)}(\lambda )],
\label{Laxpairsofconstrainedflows}\ee 
where $L^{(1)}(\la )$, $U$, and $V^{(1)}(\la )$ are given by 
(\ref{defofL^{(1)}}), (\ref{defofL^{(1)}c}), 
(\ref{defofU(f,lambda)}) and (\ref{defofV^{(1)}(f,lambda)}),
if (\ref{spatialconstrainedflowin1+1}) and 
(\ref{temporalconstrainedflowin1+1}) hold,
respectively.
\end{theorem}

{\it Proof:}  
A direct calculation can show the Hamiltonian structures of 
the spatial constrained flow (\ref{spatialconstrainedflowin1+1})
and the temporal constrained flow (\ref{temporalconstrainedflowin1+1})
with $H_1^x$ and $H_1^t$ defined by (\ref{H^x_1}) and (\ref{H^t_1}).
Let us then check the Lax representations. By using 
(\ref{spatialconstrainedflowin1+1}), 
we can compute that  
\begin{eqnarray}  && 
(L^{(1)}(\la ))_x= \sum_{s =1}^N \frac {\mu _s}{\lambda -\lambda _s }
(\phi ^{(s)}_x\psi^{(s)T}+
\phi^{(s)}\psi^{(s)T}_x )
\nonumber \\
 &= &\sum_{s =1}^N \frac {\mu _s}{\lambda -\lambda _s }
\Bigl(U({\widetilde u},\lambda_s) \phi ^{(s)}\psi^{(s)T}-
\phi^{(s)} \psi^{(s)T}U({\widetilde u},\lambda_s)  \Bigr)
\nonumber \\
&= &\sum_{s =1}^N \frac {\mu _s}{\lambda -\lambda _s }
[U(\widetilde u,\lambda_s), \phi ^{(s)}\psi^{(s)T}]
\nonumber \\
&=&[U(\widetilde u,\lambda),L^{(1)}(\lambda )-C_1]-
[U_0,\sum _{s=1}^N\mu_s \phi ^{(s)}\psi^{(s)T}]
\nonumber \\ 
& = & [U(\widetilde u,\lambda),L^{(1)}(\lambda )]+[C_1, U(\widetilde u,\lambda)]-
[U_0,\sum _{s=1}^N\mu_s \phi ^{(s)}\psi^{(s)T}]\nonumber\\
&=& 
[U(\widetilde u,\lambda),L^{(1)}(\lambda )]+[C_1, U_1(\widetilde u)]-
[U_0,\sum _{s=1}^N\mu_s \phi ^{(s)}\psi^{(s)T}]
\nonumber .
 \end{eqnarray}
This implies that  
$(L^{(1)}(\lambda ))_x
=[U(\widetilde u,\lambda ),L^{(1)}(\lambda )]$
if and only if 
\[ [C_1, U_1({\widetilde u})]=
[U_0,\sum _{s=1}^N\mu_s \phi ^{(s)}\psi^{(s)T}]. 
%\label{equiveofconstaintonpotenials}
\]
The above equality 
%(\ref{equiveofconstaintonpotenials}) 
equivalently requires the constraints on the potentials 
shown in (\ref{componentofgBscofnWIEsin1+1}).
Therefore, the spatial constrained flow (\ref{spatialconstrainedflowin1+1})
has the necessary Lax representation defined as in (\ref{Laxpairsofconstrainedflows}).
%\bea &&
%(L^{(1)}(\la ))_x= \sum_{s =1}^N \frac {\mu _s}{\lambda -\lambda _s }(\phi _{is ,x}\psi_{js }+
%\phi_{is }\psi_{js ,x} )_{n\times n}
%\nonumber \\
% &= &\sum_{s =1}^N \frac {\mu _s}{\lambda -\lambda _s }
%\Bigl( (\la _s \al _i \phi_{is } +\sum_{l\ne i} \widetilde u_{il} \phi_{ls })\psi_{js }
%-\phi_{is } (\la _s \al _j \psi_{js } +\sum_{l\ne j} \widetilde u_{lj} \psi_{ls })
% \Bigr)_{n\times n}
%\nonumber \\
% &=& \sum_{s =1}^N \frac {\mu _s} {\lambda -\lambda _s }
%\Bigl( \la _s(\al _i-\al _j) \phi_{is }\psi_{js } +
%\sum_{l\ne i} \widetilde u_{il} \phi_{ls }\psi_{js }
%- \sum_{l\ne j} \widetilde u_{lj}\phi_{is } \psi_{ls }
% \Bigr)_{n\times n},
%\nonumber \\ &&
% \left [U(\widetilde u, \lambda ), L^{(1)} ( \lambda ) \right ] = 
%[\lambda U_0+U_1, C_1+D_1]
%\nonumber \\
%&=& \lambda [U_0,D_1]+[U_1,C_1]+[U_1,D_1]
%\nonumber \\
%&=&\sum_{s =1}^N \frac {\mu _s}{\lambda -\lambda _s }\Bigl( 
%\la (\al _i-\al _j) \phi_{is }\psi_{js }
%\Bigr)_{n\times n}-\Bigl((\beta_i-\beta _j)\widetilde u_{ij}\Bigr)_{n\times n}
%\nonumber  \\
%&& + \sum_{s=1}^N\frac {\mu _s} {\la -\la _s}\Bigl (\sum_{l\ne i} \widetilde u_{il} \phi_{ls %}\psi_{js }
%-\sum_{l\ne j} \widetilde u_{lj} \phi_{is } \psi_{ls })
% \Bigr)_{n\times n}.\nonumber 
%\eea 
%These equalities imply that $(L^{(1)}(\lambda ))_x
%=[U(\widetilde u,\lambda ),L^{(1)}(\lambda )]$
%if (\ref{spatialconstrainedflowin1+1}) holds.
The proof of the other necessary Lax representation
$(L^{(1)}(\lambda ))_{t_1}=[V^{(1)}(\widetilde u,\lambda ),L^{(1)}(\lambda )]$
is completely similar, and thus we omit it. The proof is finished. 
$\vrule width 1mm height 3mm depth 0mm$

We remark that the Lax representations (\ref{Laxpairsofconstrainedflows}) 
are not sufficient. Namely, we can not obtain the spatial constrained flow
(\ref{spatialconstrainedflowin1+1}) or the temporal constrained flow
(\ref{temporalconstrainedflowin1+1}) 
from the corresponding Lax representation in (\ref{Laxpairsofconstrainedflows}).
This can be easily observed by considering 
a special class of solutions of (\ref{Laxpairsofconstrainedflows}).
For example, either any vector functions $\phi^{(s)}$ with $\psi^{(s)}=0$,
 $1\le s\le N$, 
or any vector functions $\psi^{(s)}$ with $\phi^{(s)}=0$, $1\le s\le N$, 
 will solve 
(\ref{Laxpairsofconstrainedflows}), but it is easy to see that 
they do not always 
solve (\ref{spatialconstrainedflowin1+1}) [or (\ref{temporalconstrainedflowin1+1})]
since $\phi^{(s)}$ and $\psi^{(s)}$, $1\le s\le N$, have to solve some ODEs resulting from
(\ref{spatialconstrainedflowin1+1}) [or (\ref{temporalconstrainedflowin1+1})].

\section{Binary symmetry  constraints %of the ${\cal N}$-wave interaction equations
in $2+1 $ dimensions}
\setcounter{equation}{0}

\subsection{$2+1$ dimensional ${\cal N}$-wave interaction equations}

Let $n$ be an arbitrary natural number strictly greater than two.
Similar to the case of
the 1+1 dimensional ${\cal N}$-wave interaction equations, 
let us begin with the Lax system 
\begin{equation}
   F_y=JF_x+PF,\ 
   F_t=KF_x+QF,\ F=(f_1,\cdots,f_n)^T   \label{eq:lp_nw12}
\end{equation}
in $2+1$ dimensions. Here it is assumed that 
\be J=\diag(J_1,\cdots,J_n),\ K=\diag(K_1,\cdots,K_n),\ J_i\ne J_j,\ K_i\ne K_j,\ 1\le 
i\ne j\le n \ee 
are two constant diagonal matrices, 
and $P$ and $Q$ are two $n\times n$ off-diagonal potential matrices
\be  P=P(x,y,t)=(p_{ij})_{n\times n},\  Q=Q(x,y,t)=(q_{ij})_{n\times n}. \ee
The compatability condition $F_{yt}=F_{ty}$ of the Lax system (\ref{eq:lp_nw12}) reads as 
\begin{equation}
   [J,Q]=[K,P],\ 
   P_t-Q_y+[P,Q]+JQ_x-KP_x=0,
   \label{eq:eq_nw12}
\end{equation}
which is called the $2+1$ dimensional ${\cal N}$-wave interaction
equations \cite{FokasA-JMP1984}. 
The equation  $[J,Q]=[K,P]$ tells us that $Q$ can be represented 
by $P$ and vice versa, and so practically, we have just one of two potential matrices to be 
solved. The adjoint system of the Lax system (\ref{eq:lp_nw12}) is given by 
\begin{equation}
   G_y=JG_x-P^TG,\ G_t=KG_x-Q^TG,\ \ G=(g_1,\cdots,g_n)^T,
\label{eq:alp_nw12}
\end{equation}
whose compatability condition $G_{yt}=G_{ty}$ still gives rise to 
the $2+1$ dimensional ${\cal N}$-wave interaction equations (\ref{eq:eq_nw12}).

We first use a symmetry constraint of the $2+1$ dimensional ${\cal N}$-wave interaction
equations (\ref{eq:eq_nw12})
to change the above problem in $2+1$ dimensions 
to three problems in $1+1$ dimensions. As made in 
\cite{bib:Zhou3wave,bib:Zhou2narb}, we introduce the spectral problems
\begin{equation}
\left\{   \begin{array}{l}
   \phi_x=\Omega^x(F,G,\lambda)\phi= (\lambda \Omega _0^x+\Omega _1^x)\phi 
         =\left(\begin{array}{cc}
    \lambda I_n &F\\ G^T &0\end{array}\right)\phi, \vspace{2mm}\\
   \phi_y=\Omega^y(P,F,G,\lambda)\phi= (\lambda \Omega _0^y+\Omega _1^y)\phi 
  = \left(\begin{array}{cc}
    \lambda J+P &JF\\ G^TJ &0\end{array}\right)\phi, \vspace{2mm}\\
   \phi_t=\Omega^t(Q,F,G,\lambda)\phi= (\lambda \Omega _0^t+\Omega _1^t)\phi 
  =\left(\begin{array}{cc}
    \lambda K+Q &KF\\ G^TK &0\end{array}\right)\phi,
   \end{array}\right.
   \label{eq:newlp_nw12}
\end{equation}
where $I_n$ is the $n$th-order identity matrix and $\phi=(\phi_1,\cdots ,\phi_n,\phi_{n+1})^T.$
The new extended potentials in the above spectral systems consist of not only the original
potentials, $P$ and $Q$, but also the solutions of the Lax
system and the adjoint Lax system, $F$ and $G$. 
The compatability conditions $\phi_{xy}=\phi_{yx}$,
$\phi_{xt}=\phi_{tx}$, and $\phi_{yt}=\phi_{yt}$ give rise to
the $2+1$ dimensional ${\cal N}$-wave interaction equations (\ref{eq:eq_nw12}),
the original Lax system (\ref{eq:lp_nw12}) and its adjoint system (\ref{eq:alp_nw12}),
and the nonlinear symmetry constraint of (\ref{eq:eq_nw12}):
\begin{equation}
   P_x=[FG^T,J],\ Q_x=[FG^T,K]\label{firstscofeq:eq_nw12}.
\end{equation}
It is easy to check that $(\delta P,\delta Q)=([FG^T,J],[FG^T,K])$ satisfies 
the linearized system of the $2+1$ dimensional ${\cal N}$-wave interaction
equations (\ref{eq:eq_nw12}):
\be 
\bigl[J,\delta Q \bigr]=[K,\delta P ],\ (\delta P)_t-(\delta Q) _y+[\delta P, Q]
+[P,\delta Q]
 +J(\delta Q)_x-K(\delta P)_x=0 , \label{lsofNWIEs2+1}\ee
when $F$ and $G$ solve  
the Lax system (\ref{eq:lp_nw12}) and the adjoint Lax system (\ref{eq:alp_nw12}),
respectively.
Therefore, (\ref{firstscofeq:eq_nw12}) is really a symmetry constraint of 
the $2+1$ dimensional ${\cal N}$-wave interaction
equations (\ref{eq:eq_nw12}),
since both sides of (\ref{firstscofeq:eq_nw12}) are symmetries of (\ref{eq:eq_nw12}). 
Now we see that the original problem in $2+1$ dimensions
is transformed into three problems in $1+1$ dimensions. The spectral problems 
(\ref{eq:newlp_nw12}) are our starting point to make a link of 
the $2+1$ dimensional ${\cal N}$-wave interaction equations 
(\ref{eq:eq_nw12})
to finite-dimensional integrable systems. 

\subsection{Binary symmetry constraints in $2+1$ dimensional case}

Let us start from 
the spectral problems in (\ref{eq:newlp_nw12}), which are similar to those for the
$1+1$ dimensional ${\cal N}$-wave interaction equations (\ref{nWIEsin1+1}). 
The main difference is that the coefficient matrix of $\lambda$ in the $x$-part of
the spectral problems (\ref{eq:newlp_nw12}) is 
\begin{equation}
\Omega _0^x=   \diag(\underbrace{1,\cdots,1}_n,0),
\end{equation}
whose diagonal entries are not distinct. However, the $y$-part of the spectral problems
(\ref{eq:newlp_nw12}) has 
the same property as the spectral problem (\ref{spofnWIEsin1+1}) in $1+1$ dimensions.
Therefore, we use the $y$-part of the spectral problems
(\ref{eq:newlp_nw12}) to compute the variational derivatives of $\lambda $:
 \[ \ba {l}\D \frac {\delta \lambda }{\delta p_{ij}}=E^{-1}
\psi^T\D \frac{\partial \Omega^y}{\partial p_{ij}} \phi=
E^{-1}\phi_i\psi_j, \ 
\D \frac {\delta \lambda }{\delta q_{ij}}=E^{-1}
\psi^T\D \frac{\partial \Omega^y}{\partial q_{ij}} \phi=
E^{-1}\D \frac {J_i-J_j}{K_i-K_j}\phi_i\psi_j,\ 1\le i\ne j\le n, \vspace{2mm} \\
 \D \frac {\delta \lambda }{\delta f_{i}}=E^{-1}
\psi^T\D \frac{\partial \Omega^y}{\partial f_{i}} \phi=
E^{-1}J_i\phi_{n+1}\psi_i, \ 
\D \frac {\delta \lambda }{\delta g_{i}}=E^{-1}
\psi^T\D \frac{\partial \Omega^y}{\partial g_{i}} \phi=
E^{-1}J_i\phi_{i}\psi_{n+1}, \ 1\le i\le n,
\ea   
\]
where $E$ is the normalized constant,
and $\psi=(\psi_1,\cdots ,\psi_n,\psi_{n+1})^T$ is an adjoint eigenfunction of 
the adjoint spectral problems 
\begin{equation}\left\{
   \begin{array}{l}
   \psi_x=-(\Omega^x(F,G,\lambda))^T\psi= -(\lambda (\Omega _0^x)^T+(\Omega _1^x)^T)\psi 
         =-\left(\begin{array}{cc}
    \lambda I_n &G\\ F^T &0\end{array}\right)\psi, \vspace{2mm}\\
   \psi_y=-(\Omega^y(P,F,G,\lambda))\psi= -(\lambda (\Omega _0^y)+(\Omega _1^y)^T)\psi =
-\left(\begin{array}{cc}
    \lambda J+P^T &JG\\ F^TJ &0\end{array}\right)\psi, \vspace{2mm}\\
   \psi_t=-(\Omega^t(Q,F,G,\lambda))^T\psi= -(\lambda (\Omega _0^t)^T+(\Omega _1^t)^T)\psi 
   =-\left(\begin{array}{cc}
    \lambda K+Q^T &KG\\ F^TK &0\end{array}\right)\psi.
   \end{array}\right.
   \label{eq:anewlp_nw12}
\end{equation}
These variational derivatives of $\lambda $ give us 
a conserved covariant and also a clue to compute a required symmetry, 
%of (\ref{eq:eq_nw12}), (\ref{eq:lp_nw12}) and (\ref{eq:alp_nw12}),
expressed in terms of eigenfunctions and adjoint eigenfunctions.

As in the $1+1$ dimensional case, 
upon introducing $N$ distinct eigenvalues $\lambda _1,\lambda _2,\cdots ,\lambda _N$, we have 
\begin{equation}
   \phi_x^{(s)}=\Omega^x(u,\lambda _s)\phi^{(s)},\ 
   \phi_y^{(s)}=\Omega^y(u,\lambda _s)\phi^{(s)},\ 
   \phi_t^{(s)}=\Omega^t(u,\lambda _s)\phi^{(s)},\ 1\le s\le N,
   \label{eq:Phieq}
\end{equation}
and  
\begin{equation}
   \psi_x^{(s)}=-(\Omega^x)^T(u,\lambda _s)\psi^{(s)},\ 
   \psi_y^{(s)}=-(\Omega^y)^T(u,\lambda _s)\psi^{(s)},\ 
   \psi_t^{(s)}=-(\Omega^t)^T(u,\lambda _s)\psi^{(s)},\ 1\le s\le N,
   \label{eq:Psieq}
\end{equation}
where $\phi^{(s)}$ and $\psi^{(s)}$ are $n+1$ dimensional vector functions: 
\begin{equation}
   \phi^{(s)}=(\phi_{1s},\cdots,\phi_{ns},\phi_{n+1,s})^T,\ 
   \psi^{(s)}=(\psi_{1s},\cdots,\psi_{ns},\psi_{n+1,s})^T,\ 1\le s\le N.
\end{equation}
To carry out binary nonlinearization,  
we need to construct two special symmetries, the one of which is a Lie point symmetry, and 
the other of which is not a Lie point, contact or Lie B\"acklund symmetry, but generated from 
(\ref{eq:Phieq}) and (\ref{eq:Psieq}).
Let us choose a set of $n+1$ arbitrary distinct constants
$\delta _1,\cdots, \delta _n,\delta _{n+1},$
and set 
\be \Delta =\diag (\delta _1,\cdots,\delta_n). \label{defofDelta}\ee
Similar to the $1+1$ dimensional case, it can be directly shown that  
\be 
(\delta P,\delta Q, \delta F, \delta G)= ([\Delta ,P],[\Delta ,Q],\Delta F-\delta_{n+1}F,
\Delta G-\delta _{n+1}G)
\ee 
and 
\be \left \{ \ba {l}
\delta p_{ij}=(J_i-J_j)\langle \Phi_i,B\Psi_j\rangle ,\ 
\delta q_{ij}=(K_i-K_j)\langle \Phi_i,B\Psi_j\rangle ,\ 1\le i\ne j\le n,
\vspace{2mm}\\
\delta f_i= \langle \Phi_i,B\Psi_{n+1}\rangle ,\ 
\delta g_i= \langle \Phi_{n+1},B\Psi_{i}\rangle ,\ 
1\le i\le n  ,\ea \right. \ee 
are two symmetries of 
the equations (\ref{eq:eq_nw12}), 
(\ref{eq:lp_nw12}) and (\ref{eq:alp_nw12}).
%$2+1$ dimensional ${\cal N}$-wave interaction equations (\ref{eq:eq_nw12}).
That is to say that they 
satisfy the linearized system of the equations (\ref{eq:eq_nw12}), 
(\ref{eq:lp_nw12}) and (\ref{eq:alp_nw12}):
the first subsystem (\ref{lsofNWIEs2+1})
and the second subsystem 
\be  
\ba {l}( \delta F )_y=J(\delta F) _x+(\delta P) F+P\delta F,\ (\delta F) _t=K(\delta F) _x
+(\delta Q) F+Q \delta F ,
\vspace{2mm}\\
(\delta G) _y=J(\delta G)_x-(\delta P)^TG-P^T\delta G,\ (\delta G)_t=K(\delta G) _x-
(\delta Q) ^TG-Q^T\delta G,
\ea \ee 
for all solutions $(P,Q,F,G)$ of (\ref{eq:eq_nw12}), 
(\ref{eq:lp_nw12}) and (\ref{eq:alp_nw12}).
Here we remind that \[
B=\diag(\mu_1,\cdots,\mu _N)^T\]
 is defined by 
(\ref{defofAB}), $\langle \cdot,\cdot\rangle$ denotes 
the standard inner product of $\R ^{N}$, and $\Phi_i$ and $\Psi_i$ are 
similarly defined as
\begin{equation}
   \Phi_i=(\phi_{i1},\phi_{i2},\cdots,\phi_{iN})^T,\ 
   \Psi_i=(\psi_{i1},\psi_{i2},\cdots,\psi_{iN})^T,\ 1\le i\le n+1.
\label{defofPhi_iandPsi_i2+1}
\end{equation}

Now a binary Bargmann symmetry constraint
of (\ref{eq:eq_nw12}), 
(\ref{eq:lp_nw12}) and (\ref{eq:alp_nw12})
can be taken as
\bea  %\left \{ \ba {l}
&& ([\Delta ,P])_{ij}=(J_i-J_j)\langle \Phi_i,B\Psi_j\rangle ,\ 
([\Delta, Q])_{ij}=(K_i-K_j)\langle \Phi_i,B\Psi_j\rangle ,\ 1\le i\ne j\le n,
\label{gB1scofnWIEsin2+1}\qquad \quad 
\\ &&
(\Delta F-\delta_{n+1}F)_i= \langle \Phi_i,B\Psi_{n+1}\rangle ,\ 
(\Delta G-\delta_{n+1}G)_i= \langle \Phi_{n+1},B\Psi_{i}\rangle ,\ 
1\le i\le n  .%\ea \right. 
\label{gB2scofnWIEsin2+1}
\eea 
This symmetry constraint gives us the following choice for 
the constraints on the extended potentials
\bea
   && p_{ij}= \widetilde p_{ij}:=\frac{J_i-J_j}{\delta _i-\delta _j}
    \langle\Phi_i,B\Psi_j\rangle,\  
q_{ij}=\widetilde q_{ij}:=\frac{K_i-K_j}{\delta _i-\delta _j}
    \langle\Phi_i,B\Psi_j\rangle, \  1\le i\ne j\le n,
   \label{eq:2+1constr1}\\
      && f_i=\widetilde f_i:=\frac 1 {\delta _i-\delta _{n+1}}\langle\Phi_i,B\Psi_{n+1}\rangle
    ,\ g_i=\widetilde g_i:=\frac 1 {\delta _i-\delta _{n+1}}\langle\Phi_{n+1},B\Psi_i\rangle
    ,\  1\le i\le n.\label{eq:2+1constr2} \qquad \qquad 
\eea
One can express 
the above symmetry constraint in another way.
Actually, it can be proved that 
\[  
(\delta P,\delta Q)= ([\Delta ,P],[\Delta ,Q]),\]
and under the constraint (\ref{eq:2+1constr2}),
\[ 
\delta p_{ij}=(J_i-J_j)\langle \Phi_i,B\Psi_j\rangle ,\ 
\delta q_{ij}=(K_i-K_j)\langle \Phi_i,B\Psi_j\rangle ,\ 1\le i\ne j\le n,\] 
are two symmetries of the $2+1$ dimensional $\cal N$-wave
interaction equations (\ref{eq:eq_nw12}).

Now plug the above expressions for the extended potentials, (\ref{eq:2+1constr1}) and 
(\ref{eq:2+1constr2}),
into the spectral problems (\ref{eq:newlp_nw12}) and the adjoint spectral problems
(\ref{eq:anewlp_nw12}), and then we get the constrained flows
\begin{eqnarray}&&
   \phi^{(s)}_x=\Omega^x(\widetilde F,\widetilde G,\lambda_s)\phi^{(s)},\
\psi^{(s)}_x=-(\Omega^x(\widetilde F,\widetilde G,\lambda_s))^T\psi^{(s)},\label{cf1of2+1case}
\\ &&
   \phi^{(s)}_y=\Omega^y(\widetilde P,\widetilde F,\widetilde G,\lambda_s)\phi^{(s)},\ 
\psi^{(s)}_y=-(\Omega^y(\widetilde P,\widetilde F,\widetilde 
G,\lambda _s))^T\psi^{(s)},\label{cf2of2+1case}
\\
&&   \phi^{(s)}_t=\Omega^t(\widetilde Q,\widetilde F,\widetilde G,\lambda_s)\phi^{(s)},\ 
 \psi^{(s)}_t=-(\Omega^t(\widetilde Q,\widetilde F,\widetilde G,\lambda _s))^T\psi^{(s)},
   \label{cf3of2+1case}
\end{eqnarray}
where
\begin{equation}
   \widetilde P=(\widetilde p_{ij})_{n\times n},\ 
   \widetilde Q=(\widetilde q_{ij})_{n\times n},\ 
   \widetilde F=(\widetilde f_1,\cdots,\widetilde f_n)^T,\ 
   \widetilde G=(\widetilde g_1,\cdots,\widetilde g_n)^T.
   \label{eq:constrPQFG}
\end{equation}
All these three constrained flows are 
systems of ordinary differential equations of $\phi_{is}$ and $\psi_{is}$,
$1\le i\le n+1, 1\le s\le N$.

We introduce the symplectic structure 
\be \omega ^2= \sum_{i=1}^{n+1} B d\Phi _i\wedge d\Psi _i=
\sum_{i=1}^{n+1}\sum_{s=1}^N \mu _s  d\phi_{is}\wedge d\psi _{is}
\label{eq:symplecticformin2+1}
\ee 
over $\R ^{2(n+1)N}$.
The corresponding Poisson bracket 
and the corresponding Hamiltonian form with the Hamiltonian $H$ and the evolution variable $t$
 are similarly taken as 
\bea && \{f,g\}=\sum_{i=1}^{n+1}
(\langle \frac {\part f}{\part \Psi_i},
B^{-1}\frac {\part g}{\part \Phi_i} \rangle -
\langle \frac {\part f}{\part \Phi_i},
B^{-1}\frac {\part g}{\part \Psi_i} \rangle  )
,\ f,g\in C^\infty (\R ^{2(n+1)N}), \label{eq:Poissonbracketin2+1case}
\\
 &&
\Phi_{it}=\{\Phi_{i},H\}=-B^{-1}\frac {\part H}{\part \Psi _i},\ \Psi_{it}=
\{\Psi_{i},H\}=
B^{-1}\frac {\part H}{\part \Phi_i},\ 1\le i\le n+1.
\eea 
Similar to Theorem~\ref{thm:Hamiltoniansin1+1}, 
we have

\begin{theorem}\label{thm:Hamiltoniansin2+1}
Under the symplectic structure 
(\ref{eq:symplecticformin2+1}),
 three
constrained flows (\ref{cf1of2+1case}), (\ref{cf2of2+1case}) and (\ref{cf3of2+1case}) are Hamiltonian systems with the evolution variables $x$, $y$ and $t$, 
and the Hamiltonians 
\bea
   H_2^x=&&-\sum_{k=1}^n\langle A\Phi_k,B\Psi_k\rangle
  -\sum_{k=1}^n\frac 1{\delta _k-\delta _{n+1}}\langle\Phi_k,B\Psi_{n+1}\rangle
    \langle\Phi_{n+1},B\Psi_k\rangle,\label{defofH_2^{x}}\\
   H_2^y=&&-\sum_{k=1}^nJ_k\langle A\Phi_k,B\Psi_k\rangle
   -\sum_{1\le k<l\le n}\frac{J_k-J_l}{\delta _k-\delta _l}
   \langle\Phi_k,B\Psi_l\rangle\langle\Phi_l,B\Psi_k\rangle \nonumber \\
   &&-\sum_{k=1}^n\frac{J_k}{\delta_k-\delta _{n+1}}\langle\Phi_k,B\Psi_{n+1}\rangle
    \langle\Phi_{n+1},B\Psi_k\rangle,\label{defofH_2^{y}}\\
   H_2^t=&&-\sum_{k=1}^nK_k\langle A\Phi_k,B\Psi_k\rangle
   -\sum_{1\le k<l\le n}\frac{K_k-K_l}{\delta _k-\delta _l}
   \langle\Phi_k,B\Psi_l\rangle\langle\Phi_l,B\Psi_k\rangle \nonumber\\
   &&-\sum_{k=1}^n \frac{K_k}{\delta_k-\delta _{n+1}}\langle\Phi_k,B\Psi_{n+1}\rangle
    \langle\Phi_{n+1},B\Psi_k\rangle,\label{defofH_2^{t}}
\eea 
respectively, where $A$ and $B$ are defined by 
(\ref{defofAB}), $\Phi_i$ and $\Psi_i$, $1\le i\le n+1$, are 
defined by (\ref{defofPhi_iandPsi_i2+1}).
Moreover, they possess the necessary Lax representations
\bea &&
 (L^{(2)}(\lambda ))_x=[\Omega^x(\widetilde F,\widetilde G,\lambda),L^{(2)}(\lambda ) ],
\label{xlrofcf2+1} \\
 &&(L^{(2)}(\lambda ))_y=[\Omega^y(\widetilde P,\widetilde F,\widetilde G,\lambda),L^{(2)}(\lambda ) ],\label{ylrofcf2+1}
\\ && 
 (L^{(2)}(\lambda ))_t=[\Omega^t(\widetilde Q,\widetilde F,\widetilde G,\lambda),
L^{(2)}(\lambda ) ],\label{tlrofcf2+1}
\eea 
respectively, 
where $\widetilde P$, $\widetilde Q$, $\widetilde F$ and $\widetilde G$ are given by 
(\ref{eq:constrPQFG}), 
(\ref{eq:2+1constr1}) and (\ref{eq:2+1constr2}), and $L^{(2)}(\lambda )$ is defined by 
\be \left\{
\ba {l}
L^{(2)}(\lambda )= C_2+D_2(\lambda), \ C_2=\diag (\Delta,\delta _{n+1})=\diag (\delta _1,\cdots,\delta_n,\delta_{n+1}),\vspace{2mm}\\
D_2=(D^{(2)}_{ij})_{n+1,n+1},\ D^{(2)}_{ij}=
\D \sum_{s=1}^N\frac {\mu _s}{\lambda -\lambda _s}\phi_{is}
\psi_{js},\ 1\le i,j\le n+1. \ea \right.
\ee 
\end{theorem}

{\it Proof:}
It can be verified by a direct calculation that  
all three constrained flows (\ref{cf1of2+1case}), 
(\ref{cf2of2+1case}) and (\ref{cf3of2+1case})
have the Hamiltonian structures under the symplectic structure (\ref{eq:symplecticformin2+1})
with the Hamiltonian functions 
$H_2^x$, $H_2^y$ and $H_2^t$ shown in (\ref{defofH_2^{x}}), (\ref{defofH_2^{y}}) and (\ref{defofH_2^{t}}).
Let us now check three Lax representations (\ref{xlrofcf2+1}), (\ref{ylrofcf2+1}) and (\ref{tlrofcf2+1}). Since the proofs are similar 
for all three cases, we just show the second case, i.e., the Lax representation
of the constrained flow (\ref{cf2of2+1case}). 
By using (\ref{cf2of2+1case}), 
we can compute that  
\begin{eqnarray}  && 
(L^{(2)}(\la ))_y= \sum_{s =1}^N \frac {\mu _s}{\lambda -\lambda _s }
(\phi ^{(s)}_y\psi^{(s)T}+
\phi^{(s)}\psi^{(s)T}_y )
\nonumber \\
 &= &\sum_{s =1}^N \frac {\mu _s}{\lambda -\lambda _s }
\Bigl(\Omega^y(\widetilde P,{\widetilde F},\widetilde G,\lambda_s) \phi ^{(s)}\psi^{(s)T}-
\phi^{(s)} \psi^{(s)T}
\Omega^y(\widetilde P,{\widetilde F},\widetilde G,\lambda_s)   \Bigr)
\nonumber \\
&= &\sum_{s =1}^N \frac {\mu _s}{\lambda -\lambda _s }
[\Omega^y(\widetilde P,{\widetilde F},\widetilde G,\lambda_s) , \phi ^{(s)}\psi^{(s)T}]
\nonumber \\
&=&\sum _{s=1}^N\frac {\mu_s} {\lambda -\lambda _s} 
( [\Omega^y(\widetilde P,{\widetilde F},\widetilde G,\lambda)
, \phi ^{(s)}\psi^{(s)T}]-
[\Omega^y(\widetilde P,{\widetilde F},\widetilde G,\lambda)
-\Omega^y(\widetilde P,{\widetilde F},\widetilde G,\lambda_s)
, \phi ^{(s)}\psi^{(s)T}])
\nonumber \\
&=&[
\Omega^y(\widetilde P,{\widetilde F},\widetilde G,\lambda)
,L^{(2)}(\lambda )-C_2]
%-[\Omega^y(\widetilde P,{\widetilde F},\widetilde G,\lambda) ,C_2]
-[\Omega^y_0,\sum _{s=1}^N\mu_s \phi ^{(s)}\psi^{(s)T}]
\nonumber \\ 
&=&[\Omega^y(\widetilde P,{\widetilde F},\widetilde G,\lambda)
,L^{(2)}(\lambda )]-[\Omega^y_1(\widetilde P,{\widetilde F},\widetilde G)
 ,C_2]-
[\Omega^y_0,\sum _{s=1}^N\mu_s \phi ^{(s)}\psi^{(s)T}]
.\nonumber
 \end{eqnarray}
Therefore, it follows that   
$(L^{(2)}(\lambda ))_y
=[\Omega^y(\widetilde P,{\widetilde F},\widetilde G,\lambda)
,L^{(2)}(\lambda )]$
if and only if 
\[
[C_2,\Omega^y_1(\widetilde P,{\widetilde F},\widetilde G) ]=
[\Omega^y_0,\sum _{s=1}^N\mu_s \phi ^{(s)}\psi^{(s)T}].
\]
This equality 
%(\ref{equiveofconstaintonpotenials}) 
equivalently requires the nonlinear constraints on the potentials 
defined by (\ref{eq:2+1constr1}) and (\ref{eq:2+1constr2}).
Therefore, the constrained flow (\ref{cf2of2+1case})
has the necessary Lax representation shown in (\ref{ylrofcf2+1}).
The proof is finished. 
$\vrule width 1mm height 3mm depth 0mm$

We also remark that the Lax representations (\ref{xlrofcf2+1}), 
(\ref{ylrofcf2+1}) 
and (\ref{tlrofcf2+1}) 
are not sufficient to generate 
the corresponding constrained flows defined by (\ref{cf1of2+1case}), (\ref{cf2of2+1case})
and (\ref{cf3of2+1case}), since the Gateaux derivative operators of 
the Lax operators $\Omega^x$, $\Omega^y$ and $\Omega^t$ 
given in (\ref{cf1of2+1case}), (\ref{cf2of2+1case})
and (\ref{cf3of2+1case}) are not injective. However, it will 
be shown that they are good enough in generating
integrals of motion of the constrained flows. 

\section{An involutive and functionally independent system of polynomial functions
%from an arbitrary-order matrix Lax operator
}
\setcounter{equation}{0}

Let $m$ be an arbitrary natural number.
We start from an $m$-order matrix Lax operator
\be L(\lambda )=L(\la ;c_1,\cdots,c_m ) =C+D(\la ), \label{defofL(lambda)}\ee 
with $C$ and $D(\la )$ being defined by  
\be  C=\textrm{diag}(c _1,\cdots, c _m ),\ 
D(\la )=(D_{ij}(\la ))_{m\times m},\
D_{ij}(\la )=\sum_{s =1}^N\frac {\mu _s}{\la -\la _s } \phi _{is}\psi_{js},\ 
1\le i,j\le m.\label{defofL(lambda)a}\ee 
Here $c _i,\ \la _s,$ and $ \mu _s$ are arbitrary constants satisfying 
\be  \prod _{s=1}^N \mu _s\ne 0,\ \la _i\ne \la _j, \ 1\le i\ne j\le N,  \ee 
and $ \phi_{is}$ and $\psi_{js}$ are pairs of 
canonical variables of the symplectic manifold $(\R ^{2m N},\omega ^2)$
with the symplectic structure 
\be \omega ^2=\sum_{i=1}^m  \sum_{s=1}^N \mu_s d\phi_{is}\wedge d\psi_{is}. \ee 
The corresponding Poisson bracket reads as
\be 
\{f,g\}= \omega ^2(Idg,Idf)=
\sum_{i=1}^m \sum_{s=1}^N \mu_s ^{-1}\Bigl( \frac {\part f}{\part \psi_{is}}\frac {\part g}{\part \phi_{is}}
- \frac {\part f}{\part \phi_{is}}\frac {\part g}{\part \psi_{is}}
\Bigr),\ f,g\in C^\infty (\R ^{2m N}).
\label{Poissonbracketofmorder}
\ee  

\subsection{${\bf r}$-matrix formulation}  

As usual, two special matrices defined by the tensor product of 
matrices are chosen as
\begin{equation}
L _1(\lambda )=L (\lambda )\otimes I_m  ,
\ L _2(\mu  )=I_m  \otimes L (\mu  ),
%\ I_m  =\textrm{diag}(\underbrace{1,\cdots,1}_m  ).
\end{equation}
where $I_m  $ is the $m$th-order identity matrix,
and 
\be  (A\otimes B)_{ij,kl}=a_{ik}b_{jl} \ 
 \textrm{if} \ A=(a_{ij})\  \textrm{and} \   B=(b_{ij}).\ee 
We want to find an $m^2\times m^2$ matrix ${\bf r}={\bf r}(\lambda,\mu )$ so that 
we have a ${\bf r}$-matrix formulation 
\cite{Semenov-Tian-Shansky-FAP1983,FaddeevT-book1987}
\begin{equation}
%\{L _{1}(\lambda),\{L _{2}(\mu )\}=
\{L (\lambda)\stackrel{\otimes}{,} L (\mu)\}
=[{\bf r}(\lambda ,\mu ),
L _{1}(\lambda)+L _{2}(\mu)],
\end{equation}
with the Poisson bracket $\{L(\la ) \stackrel{\otimes}{,} L (\mu)\}$
being defined by 
\be  (\{L (\lambda)\stackrel{\otimes}{,}L (\mu)\})_{ij,kl}
=\{L _{ik}(\lambda),L _{jl}(\mu)\}=\omega ^2(IdL_{jl}(\mu ),IdL_{ik}(\la ))
,\ 1\le i,j,k,l\le m ,\qquad \ 
\ee 
where $L=(L_{ij})_{m\times m}$ is assumed.
Let us first compute $\{L _{ij}(\lambda),L _{kl}(\mu )\}$.
When $i\ne l$ and $ j\ne k$, it is easy to obtain 
$\{L _{ij}(\lambda),L _{kl}(\mu )\}=0$. When 
$i\ne l$ and $ j= k$,
we have 
\[ \begin{array}{l}
\{L _{ij}(\lambda),L _{jl}(\mu )\}=
\displaystyle{ \sum_{s =1}^N \mu _s \frac {\phi _{is } }{\lambda -\lambda _s } 
\frac {\psi_{ls } } {\mu -\lambda _s }
}
\vspace{2mm}\\
\displaystyle{ =\sum_{s =1}^N 
\frac 1{\mu -\lambda }(\frac {\mu _s} {\lambda -\lambda _s }
-\frac {\mu _s} {\mu -\lambda _s })\phi_{is } \psi _{ls }
=\frac 1 {\mu -\lambda }
(L _{il}(\la  )-L _{il}(\mu )).
}
\end{array}
\]
Similarly, when 
$i= l$ and $ j\ne k$,
we have 
\[ 
\{L _{ij}(\lambda),L _{ki}(\mu )\}=-
\sum_{s =1}^N {\mu _s}\frac {\psi _{js } }{\lambda -\lambda _s } \frac {\phi_{ks } } 
{\mu  -\lambda _s }=\frac 1 {\mu -\lambda }
(L _{kj}(\mu   )-L _{kj}(\la   )),
\]
and when 
$i=l$ and $ j= k$,
we have 
\begin{eqnarray}&& 
\{L _{ij}(\lambda),L _{ji}(\mu )\}=
\sum_{s =1}^N {\mu _s}\frac {\phi _{is } }{\lambda -\lambda _s } \frac {\psi_{is } } 
{\mu  -\lambda _s }-
\sum_{s =1}^N {\mu _s}\frac {\psi _{js } }{\lambda -\lambda _s } \frac {\phi_{js } } 
{\mu  -\lambda _s }
 \nonumber \\ &&
=\frac 1 {\mu -\lambda }
[(L _{ii }(\lambda )-L _{ii}(\mu  ))-
(L _{jj}(\lambda )-L _{jj}(\mu  ) ) 
].\nonumber 
\end{eqnarray}
Therefore, we obtain
\begin{equation}
\{L _{ij}(\lambda),L _{kl}(\mu )\}=\left\{ 
\begin{array}{ll}
0, & \mbox{when}\ i\neq l,\ j\neq k; \vspace{2mm}\\ 
\frac 1{\mu -\lambda }(L _{kj}(\mu  )-L _{kj}(\la  )),&
\mbox{when}\ i=l,\ j\ne k;  \vspace{2mm}\\ 
\frac 1{\mu -\lambda }(L _{il}(\lambda   )-L _{il}(\mu  )),&
\mbox{when}\ i\ne l,\ j= k;  \vspace{2mm}\\ 
\frac 1{\mu -\lambda }[(L _{ii}(\lambda  )-L _{ii}(\mu ) )-
(L _{jj}(\lambda )-L _{jj}(\mu ))], &\mbox{when}\ i=l,\ j=k. 
\end{array} \right.\label{leftsideofrmatrixofNWIE}
\end{equation}
In view of this property, we claim that 
\begin{equation}
{\bf r}(\lambda ,\mu )=\frac 1 {\mu -\lambda}{\cal P},\ {\cal P}=\sum_{p,q=1}^m  E_{pq}\otimes
E_{qp}, 
\end{equation}
where $E_{pq}$ is an $m  \times m  $ matrix with the $(p,q)$ entry
being one but the others, zero.
Let us second compute that 
\begin{eqnarray}
&&([\frac 1 {\mu -\lambda}{\cal P},L _1(\lambda )+L _2(\mu )] )_{ij,kl}\nonumber \\&&
=  \frac 1 {\mu -\lambda}\{[{\cal P},L _1(\lambda )]+[{\cal P},L _2(\mu )]\}_{ij,kl}\nonumber \\
&&=\frac 1 {\mu - \lambda }\sum _{p,q=1}^m   ([E_{pq},L (\lambda )]\otimes E_{qp}
+E_{qp}\otimes [E_{pq},L (\mu )])_{ij,kl}\nonumber \\&& 
=\frac 1 {\mu -\lambda  }\sum _{p,q=1}^m   ([E_{pq},L (\lambda )]_{ik} (E_{qp})_{jl}
+(E_{qp})_{ik} [E_{pq},L (\mu )]_{jl}\nonumber \\&& 
=\frac 1 {\mu -\lambda } ([E_{lj},L (\lambda )]_{ik} 
+[E_{ki},L (\mu )]_{jl},\nonumber 
\end{eqnarray}
where we have used $(A\otimes B)(A'\otimes B')=(AA')\otimes (BB')$.
Further noting that 
\[ [E_{pq},L ]=E_{pq}L -L E_{pq}= \stackrel{
\left.  \begin{array}{c}\  \vspace{2mm} \end{array}\right. 
\left. \begin{array}{ccccc} & & \ \ \ q\textrm{th}& &  \end{array}\right. }
{\left. \begin{array}{c}
 \ \vspace{2mm}\\ 
\  \vspace{4mm}\\ 
 p\textrm{th} \vspace{2mm}\\ 
\  \vspace{2mm}\\ 
\ \vspace{2mm} 
 \end{array}\right.
\left [ \begin{array}{ccccc}
 0 & \cdots & -L _{1p}&\cdots & 0  \vspace{2mm}\\ 
\vdots   &  & \vdots  &  & \vdots    \vspace{2mm}\\ 
L _{q1} & \cdots & L _{qq}-L _{pp} &\cdots & L _{qm  }  \vspace{2mm}\\ 
\vdots  &  & \vdots  &  &  \vdots   \vspace{2mm}\\ 
0 & \cdots & -L _{m  p}&\cdots & 0  
\end{array}\right]} ,
\]
we have
\begin{eqnarray}&&
([\frac 1 {\mu -\lambda}{\cal P},L _1(\lambda )+L _2(\mu )] )_{ij,kl}\nonumber \\&&
=\left\{ 
\begin{array}{ll}
0, & \mbox{when}\ i\neq l,\ j\neq k; \vspace{2mm}\\ 
\frac 1{\mu -\lambda }(L _{jk}(\lambda )-L _{jk}(\mu )),&
\mbox{when}\ i=l,\ j\ne k;  \vspace{2mm}\\ 
\frac 1{\mu -\lambda }(-L _{il}(\lambda )+L _{il}(\mu )),&
\mbox{when}\ i\ne l,\ j= k;  \vspace{2mm}\\ 
\frac 1{\mu -\lambda }[(L _{jj}(\lambda )-L _{ii}(\lambda))+
(L _{ii}(\mu )-L _{jj}(\mu ) )], &\mbox{when}\ i=l,\ j=k. 
\end{array} \right.\label{rightsideofrmatrixofNWIE}
\end{eqnarray}
Now (\ref{leftsideofrmatrixofNWIE}) and (\ref{rightsideofrmatrixofNWIE}) 
shed right on
 the following theorem.

\begin{theorem} If $L(\la )=L(\la ; c_1,\cdots, c_m)$ is defined by 
(\ref{defofL(lambda)}) and (\ref{defofL(lambda)a}), then  
the ${\bf r}$-matrix formulation 
%for the constrained flow (\ref{cfofNWIE}) of the multiple wave interaction system: 
\begin{equation}
\{L (\lambda)\stackrel{\otimes}{,}L (\mu)\}=[{\bf r}(\la ,\mu ),
L (\lambda)\otimes I_m+I_m\otimes L (\mu)],\ {\bf r}
=\frac{1}{\mu-\lambda}\sum_{i,j=1}^mE_{ij} \otimes E_{ji}
 \label{rmatrixofNWIE}
\end{equation}
holds for arbitrary constants $c_1,c_2,\cdots,c_m$.
\end{theorem}

It follows from (\ref{rmatrixofNWIE}) that 
\begin{equation}
\{L ^k (\lambda)\stackrel{\otimes}{,}L ^l (\mu)\}=[
{\bf r}^{k ,l }(\lambda ,\mu ),
L _{1}(\lambda)+L _{2}(\mu)],\ k,l\ge 1, \label{traceofL^kL^l} \end{equation}
where ${\bf r}^{k ,l }(\lambda ,\mu )$ is given by \cite{BabelonV-PLB1990}
\begin{equation}
 {\bf r}^{k ,l }(\lambda ,\mu )=
\sum_{i=1}^{k }\sum_{j=1}^{l }L _1^{k -i}(\lambda )L _2^{l -j}(\mu )
{\bf r}(\lambda ,\mu ) L _1^{i-1}(\lambda) L _2^{j-1}(\mu ). 
\end{equation} 
Since for $A=(a_{ij})_{m\times m}$ and $B=(b_{ij})_{m\times m}$ we have 
\be  
\textrm{tr}\,\{A\stackrel{\otimes}{,}B\}= 
\sum_{i,j=1}^m  \{A\stackrel{\otimes}{,}B\}_{ij,ij}
=\sum_{i,j=1}^m  \{a_{ii},b_{jj} \}=
\{\textrm{tr}\,A,\textrm{tr}\,B\},
\nonumber 
\ee
we can compute, based on (\ref{traceofL^kL^l}),
that 
\bea &&  
\{\textrm{tr}\,L ^k (\lambda) , \textrm{tr}\,L ^l  (\mu ) \}
=\textrm{tr}\, \{L ^k (\lambda)\stackrel{\otimes}{,}L ^l (\mu)\}
\nonumber \\ &&
= \textrm{tr}\,[{\bf r}^{k ,l }(\lambda ,\mu ),L _{1}(\lambda)+L _{2}(\mu)]
=0,\ k ,l \ge 1. \label{twotracesarezero}
\eea 
This will be used to generate an involutive system of functions 
defined over the symplectic manifold 
$(\R ^{2mN}, \omega ^2)$ for any natural number $m$.

\subsection{An involutive and functionally independent system}

Let us begin to construct an involutive system of polynomial functions
by expanding 
%is an independent constant of the variable $x$ with the expansion 
\be 
{\rm det}(\nu I_m -L(\lambda ))=\nu^m  -{\cal F}^{(1)}_\lambda\nu^{m  -1}+{\cal F}^{(2)}_\lambda\nu^{m  -2}+\cdots+(-1)^m  {\cal F}^{(m  )}_\lambda,\ 
\nu =\textrm{const.},
 \label{mu_-L_1} 
\ee 
where ${\cal F}^{(k)}_\la $, $1\le k\le m$, must read as 
\be 
{\cal F}^{(k )}_\lambda={\cal F}^{(k )}_\lambda(c_1,\cdots,c_m)=
{\displaystyle \sum_{1\leq j_1<j_2<\cdots<j_k\leq m  }}
\left|\matrix{L_{j_1j_1}&L_{j_1j_2}&\cdots&L_{j_1j_k}\vspace{2mm}
\cr L_{j_2j_1}&L_{j_2j_2}&\cdots&L_{j_2j_k}
\vspace{2mm}
\cr \vdots&\vdots&\ddots&\vdots
\vspace{2mm}
\cr L_{j_kj_1}&L_{j_kj_2}&\cdots&L_{j_kj_k}\cr}\right|,\ 1\le k\le m.
\label{defofcalF^i}\ee  
Here we mention once more that $ L= (L_{ij})_{m \times m } $ is assumed. 
We define bilinear functions $\stackrel{ij}{Q}_\lambda$ on $\R ^N$ 
\be 
\stackrel{ij}Q_\lambda
=\sum^N_{s=1}\mu _s
\frac{\phi_{is} \psi_{js}}
{\lambda-\lambda_s}=\sum_{l\geq 0}\langle A ^l \Phi_i,B \Psi_j\rangle\lambda^{-l-1},
\ 1\le i,j\le m,
\ee 
where $A$ and $B$ are given by (\ref{defofAB}), and 
$\Phi _i$ and $\Psi _i$ are defined as before
\begin{equation}   \Phi_i= (\phi_{i1},\phi_{i2},\cdots,\phi_{iN})^T,\ \Psi_i=
(\psi_{i1},\psi_{i2},\cdots,\psi_{iN})^T,\ 1\le i\le m .
\label{defofPhi_iandPsi_iuptom}\end{equation}  
Then we have 
\bea &&  
L_{ij}=\sum_{l\geq 0}\langle A ^{l}\Phi _i,B\Psi_j\rangle\lambda^{-l-1}=\stackrel{ij}Q_\lambda,\ 
1\leq i\ne j\leq m  ,
\nonumber \\ &&
L_{ii}=c _i+\sum_{l\geq 0}\langle A ^{l}\Phi_i,B\Psi_i\rangle\lambda^{-l-1}=c _i+\stackrel{ii}Q_\lambda,\ 1\leq i\leq m  .
\nonumber \eea
Therefore, 
the system of functions ${\cal F}^{(k)}_\la $ is transformed into 
\be 
{{\cal F}}^{(k)}_\lambda=\sum_{1\leq j_1<j_2<\cdots<j_k\leq m  }
\left|\matrix{c _{j_1}+\stackrel{j_1j_1}Q_\lambda&\stackrel{j_1j_2}
Q_\lambda&\cdots&\stackrel{j_1j_k}Q_\lambda
\vspace{2mm}
\cr \stackrel{j_2j_1}Q_\lambda&c _{j_2}+\stackrel{j_2j_2}Q_\lambda&\cdots&\stackrel{j_2j_k}Q_\lambda
\vspace{2mm}
\cr \vdots&\vdots&\ddots&\vdots
\vspace{2mm}
\cr \stackrel{j_kj_1}Q_\lambda&\stackrel{j_kj_2}Q_\lambda&\cdots&c _{j_k}+\stackrel{j_kj_k}Q_\lambda\cr}\right|,\ 1\le k\le m. \label{eq:secondformulaforcalF^k_lambda}
\ee 
A set of more concrete formulas for computing ${{\cal F}}^{(k)}_\lambda$ will be given in Appendix B.
Now we further expand ${\cal F}^{(k)}_\lambda $ as a power series of $1/\lambda $:
\be 
{\cal F}^{(k)}_\la ={\cal F}^{(k)}_\la (c_1,\cdots,c_m )
 =\sum_{l\ge 0} F_{kl}(c_1,\cdots,c_m ) \lambda ^{-l},\ 1\le k\le m .
\ee 
Based on the formulas of ${\cal F}^{(k)}_\lambda $ in Appendix~\ref{AppendixB},
it is not difficult to find that 
\bea   
&&
F_{k0}=F_{k0}(c_1,\cdots,c_m )=
\sum_{1\leq j_1<j_2<\cdots<j_k\leq m  }\prod_{p=1}^kc_{j_p},
\nonumber \\
&&
F_{kl}=F_{kl}(c_1,\cdots,c_m )=
\sum_{1\leq j_1<j_2<\cdots <j_k\leq m  } \sum_{r=1}^
{ \textrm{{\scriptsize min}}(k,l)}
\sum_{1\leq i_1<i_2<\cdots<i_r\leq k  }
\prod_{\stackrel{p=1}{p\ne i_1,i_2,\cdots,i_r}}^kc _{j_p} 
\nonumber \\
&& \times \sum_{\scriptstyle p_1+p_2+\cdots+p_r=l-r\atop
    \scriptstyle p_1,p_2,\cdots,p_r\ge 0}
   \left|\begin{array}{cccc}
    \langle A^{p_1}\Phi_{j_{i_1}},B\Psi_{j_{i_1}}\rangle
    &\langle A^{p_2}\Phi_{j_{i_2}},B\Psi_{j_{i_1}}\rangle
    &\cdots &\langle A^{p_r}\Phi_{j_{i_r}},B\Psi_{j_{i_1}}\rangle \vspace{2mm}
     \\
    \langle A^{p_1}\Phi_{j_{i_1}},B\Psi_{j_{i_2}}\rangle
    &\langle A^{p_2}\Phi_{j_{i_2}},B\Psi_{j_{i_2}}\rangle
    &\cdots &\langle A^{p_r}\Phi_{j_{i_r}},B\Psi_{j_{i_2}}\rangle\vspace{2mm}
    \\
    \vdots &\vdots &\ddots &\vdots \vspace{2mm}
     \\
    \langle A^{p_1}\Phi_{j_{i_1}},B\Psi_{j_{i_r}}\rangle
    &\langle A^{p_2} \Phi_{j_{i_2}},B\Psi_{j_{i_r}}\rangle
    &\cdots &\langle A^{p_r}\Phi_{j_{i_r}},B\Psi_{j_{i_r}}\rangle
    \end{array}\right|,\ l\ge 1, \label{defofF_{il}}\qquad\quad  
\eea 
which are all polynomials in the canonical 
variables $\phi_{is}$ and $\psi_{is},$ $1\le i\le m$, $
1\le s\le N$.
 
\begin{theorem} \label{maintheoremofintegralsofmotion1}
For all constants $c_1,c_2,\cdots,c_m$, the polynomial functions in
$\phi_{is}$ and $\psi_{is},$ $1\le i\le m$, $
1\le s\le N$:
$F_{il}(c_1,\cdots,c_m )$, $1 \le i\le m ,$ $l\ge 1$, 
defined by (\ref{defofF_{il}}),
are in involution in pair with respect to the Poisson bracket 
(\ref{Poissonbracketofmorder}).
\end{theorem}

{\it Proof:}
On the one hand, by using Newton's identities on elementary symmetric polynomials
\cite{BorweinE-book1995}
\[ \zeta  _k(\la ) -{\cal F}^{(1)}_\la \zeta  _{k-1}(\la ) 
+{\cal F}^{(2)}_\la \zeta  _{k-2}(\la )  
+\cdots + (-1)^{k-1}{\cal F}^{(k-1)}_\la \zeta  _{1}(\la )
+(-1)^kk{\cal F}^{(k)}_\la =0,
\]
where $1\le k\le m$ and 
\[ \zeta _i(\la )=
\textrm{tr}L^i(\lambda )
,\ 1\le i\le m,\]
we can have 
\be  {\cal F}^{(k)}_\la = {\cal F}^{(k)}_\la (\zeta  _1(\la ),\zeta _2(\la ),\cdots,\zeta _k(\la )),
\  1\le k\le m  .  \label{Newtonf} \ee 
Therefore, we can compute that  
\bea && \{ 
{\cal F}^{(k)}_\la ,{\cal F}^{(i)}_\mu \}= \{
{\cal F}^{(k)}_\lambda (\zeta _1(\lambda ),\zeta _2(\lambda ),
\cdots, \zeta  _k(\lambda )),{\cal F}^{(i)}_\mu  (\zeta  _1(\mu ),\zeta _2(\mu  ),
\cdots, \zeta  _i(\mu  )) \}
\nonumber \\&&
=\sum_{l=1}^k\sum_{j=1}^i
\frac{\part {\cal F}^{(k)}_\la }
{\part \zeta  _l(\la ) }\frac{{\cal F}^{(i)}_\mu }{\part \zeta  _j(\mu )}
\{\textrm{tr}L^l(\lambda ),\textrm{tr}L^j(\mu  ) \} 
=0,\  1 \le k,i \le m .
\nonumber \eea 
The last equality is a consequence of the involutivity 
of $\zeta_i(\lambda )$, $1\le i\le m$, shown in 
(\ref{twotracesarezero}).
On the other hand, we have 
\[\{{\cal F}^{(k)}_\la ,{\cal F}^{(i)}_\mu \} = \sum_{l,j\ge 0}
\{F_{kl},F_{ij}\}\la ^{-l}\mu  ^{-j}.
 \]
It follows that the polynomial functions 
$F_{il}=F_{il}(c_1,\cdots,c_m ),$ $1\le i\le m ,$ $ l\ge 1$, are in 
involution in pair with respect to the Poisson bracket
(\ref{Poissonbracketofmorder}).
$\vrule width 1mm height 3mm depth 0mm$

Let us now go on to show the functional independence of 
the polynomial functions
$F_{is}(c_1,\cdots,c_m )$, $1 \le i\le m ,\ 1\le s\le N$.

\begin{theorem} \label{maintheoremofintegralsofmotion2}
If all constants $c_1,c_2,\cdots,c_m $ are distinct, then 
the polynomial functions in 
$\phi_{is}$ and $\psi_{is},$ $1\le i\le m$, $
1\le s\le N$:
$F_{is}(c_1,\cdots,c_m )$, $1 \le i\le m ,\ 1\le s\le N$, 
defined by (\ref{defofF_{il}}),
are functionally independent over a dense open subset of 
$\R ^{2mN}$.
\end{theorem}

{\it Proof:}
Let $P_0$ be a point of $\R ^{2mN}$ satisfying 
\[\phi_{is}=\varepsilon,\ 1\le i\le m, \ 1\le s\le N,\]
where $\varepsilon$ is a small constant. Keep 
(\ref{defofF_{il}}) in mind, and then 
at this point $P_0$, we obviously have 
\bea 
   \frac{\partial F_{is_1}}{\partial \psi_{js_2}}
   &=&  \frac{\partial }{\partial \psi_{js_2}}\sum_{1\le j_1<j_2<\cdots <j_i\le m}
\sum_{q=1}^i \prod _{p=1\atop p\ne q}^i c_{j_p}
\langle A^{s_1-1}\Phi_{j_q},B\Psi_{j_q} \rangle +\textrm{O}(\varepsilon^2) \nonumber \\
&=& 
\varepsilon\sum_{1\le j_1<j_2<\cdots <j_{i-1}\le m\atop j_1,j_2,\cdots, j_{i-1}\ne j}
c_{j_1}c_{j_2}\cdots c_{j_{i-1}}\lambda_{s_2}^{s_1-1}\mu _{s_2}+\textrm{O}(\varepsilon^2), 
%\label{frac{partialF_{is_1}}{partialpsi_{js_2}}}
\eea
where $1\le i,j\le m,$ $ 1\le s_1,s_2\le N$.
In the above computation, only the term 
with $r=1$ in the expression (\ref{defofF_{il}}) of $F_{is}$ 
contributes to the first-order term of $\varepsilon$. 
Let the matrix $\Theta _N$ be defined by 
\[\Theta_N =(\Theta ^{(N)}_{ij})_{N\times N} ,\ \Theta^{(N)}_{ij}
=\lambda _{i}^{j-1}\mu _i,\ 
1\le i,j\le N, \]
whose determinant is easily found to be  
\[\det (\Theta_N) =\prod _{i=1}^N\mu _i \prod _{1\le i<j\le N}(\lambda _j-\lambda _i). \]
Then at the point $P_0$, the Jacobian of the functions 
$F_{is_1}$ with respect to $\psi _{js_2}$ can be computed as follows
\bea 
&&   \frac{\partial(F_{11},\cdots,F_{1N},
    F_{21},\cdots,F_{2N},\cdots,
    F_{m1},\cdots,F_{mN})}
   {\partial(\psi_{11},\cdots,\psi_{1N},
    \psi_{21},\cdots,\psi_{2N},\cdots,
    \psi_{m1},\cdots,\psi_{mN})} \nonumber \\
  & =& \varepsilon^{mN}
\left |
\ba {ccccc} \Theta_N &\D  \sum_{i=2}^m c_i \Theta_N  & \D 
\sum_{2\le i<j\le m}c_ic_j \Theta_N 
%&\D \sum_{2\le i<j<k\le m}c_ic_jc_k \Theta_N 
&\cdots & 
\D \prod _{i=2}^mc_i \Theta_N \vspace{2mm}\\
\Theta_N  & \D \sum_{i=1\atop i\ne 2}^m c_i \Theta_N & \D 
\sum_{1\le i<j\le m\atop i,j\ne 2}c_ic_j \Theta_N 
%&\D \sum_{1\le i<j<k\le m\atop i,j,k\ne 2}c_ic_jc_k \Theta_N 
&\cdots & 
\D \prod _{i=1\atop i\ne 2}^mc_i\Theta_N  \vspace{2mm}\\
\vdots & \vdots & \vdots &\ddots & \vdots \vspace{2mm}\\
\Theta_N  &\D  \sum_{i=1}^{m-1} c_i\Theta_N  & 
\D \sum_{1\le i<j\le m-1}c_ic_j\Theta_N  
%&\D \sum_{1\le i<j<k\le m-1}c_ic_jc_k \Theta_N 
&\cdots & \D \prod _{i=1}^{m-1}c_i \Theta_N 
 \ea \right|
+\textrm{O}(\varepsilon ^{mN+1})\nonumber \\
&= & \varepsilon^{mN}
\det (\Omega  _m\otimes \Theta_N )
+\textrm{O}(\varepsilon ^{mN+1})\nonumber \\
&=& \varepsilon^{mN}
(\det (\Omega  _m))^N(\textrm{det}( \Theta_N))^m 
+\textrm{O}(\varepsilon ^{mN+1})\nonumber \\
&=& \varepsilon^{mN}
\prod_{1\le i<j\le m}(c_i-c_j)^N \prod _{i=1}^N\mu _i\prod _{1\le i<j\le N}
(\lambda _j-\lambda _i)^m
+\textrm{O}(\varepsilon ^{mN+1}),\nonumber 
\eea 
where we have used the determinant property of the tensor product of matrices and 
the determinant result of the matrix $\Omega  _m$ in Appendix~\ref{app:detofOmega_m}.
This allows to conclude that if the constants $c_1,c_2,\cdots, c_m$ are distinct,
the above Jacobian is not zero at $P_0$ when $\varepsilon \ne 0$ is small enough. Since the Jacobian is a polynomial function of 
$\phi_{is}$ and $\psi_{is}$, $1\le i\le m$, $1\le s\le N$,
it is not zero over a dense open subset of $\R^{2mN}$. 
Therefore, the functions $F_{is},\ 1\le i\le m, \ 1\le s\le N$,  
are functionally independent over that dense open subset of $\R^{2mN}$.
The proof is complete. 
$\vrule width 1mm height 3mm depth 0mm$

\subsection{An alternative involutive system to the $F_{is}$'s}
\label{subsec:alternativeisE_{is}}

We would like to express the involutive system of the polynomial functions 
$F_{is}$ in another way, and so we introduce 
\begin{equation}
  \left\{ \begin{array}{l}
   s_0(v_1,\cdots,v_m)=1,\vspace{2mm}\\
   \displaystyle s_k(v_1,\cdots,v_m)
    =\sum_{1\le j_1<j_2<\cdots<j_k\le m}v_{j_1}\cdots v_{j_k},\ 1\le k\le m,
\vspace{2mm}\\
s_k(v_1,\cdots,v_m)=0, \   \textrm{when}\  k\ge m+1\ \textrm{or}\  k\le -1,
   \end{array}\right.
\label{eq:s_j}
\end{equation}
where $v_1, v_2, \cdots, v_m$ are $m$ numbers.
Obviously, for $m\ge 2$, we have the following relation
\begin{equation}
   s_k(v_1,\cdots,v_m)=v_ms_{k-1}(v_1,\cdots,v_{m-1})+s_k(v_1,\cdots,v_{m-1}),
\ k\in \Z .
   \label{eq:sdecomp}
\end{equation}
Let us now define 
\begin{equation}
   E_{1l}=F_{1l},\ E_{il}=(-1)^{i+1}F_{il}
   +\sum_{j=1}^{i-1}(-1)^{j+1}\sj{j}{m} E_{i-j,l}, \ i\ge 2,\ l\ge 1.
\label{defofE_{il}}
\end{equation}
From (\ref{defofE_{il}}),  we can have 
\be  F_{il}= \sum_{j=0}^{i-1} (-1)^{i-j+1}  
\sj{j}{m}E_{i-j,l},\ i,l\ge 1.
\ee 
Therefore, by Proposition~\ref{prop:sum2} in Appendix~\ref{appx:twoidentitiesforsp}, we obtain
\begin{equation}
   \begin{array}{rl}
   E_{il}=&E_{il}(c_1,\cdots,c_m)=\D\sum_{r=1}^{{\scriptstyle  {\textrm{\scriptsize min}}}(i,l)}(-1)^{r+1}
   \sum_{1\le j_1<j_2<\cdots<j_r\le m}
   \sum_{\scriptstyle l_1+l_2+\cdots+l_r=i-r\atop
    \scriptstyle l_1,l_2,\cdots,l_r\ge 0}
    c_{j_1}^{l_1}c_{j_2}^{l_2}\cdots c_{j_r}^{l_r}\vspace{2mm}\\
   &\D\times
   \sum_{\scriptstyle p_1+p_2+\cdots+p_r=l-r\atop
    \scriptstyle p_1,p_2,\cdots,p_r\ge 0}
   \left|\begin{array}{cccc}
    \langle \Lambda^{p_1}\Phi_{j_1},B\Psi_{j_1}\rangle
    &\langle \Lambda^{p_2}\Phi_{j_2},B\Psi_{j_1}\rangle
    &\cdots &\langle \Lambda^{p_r}\Phi_{j_r},B\Psi_{j_1}\rangle \vspace{2mm}\\
    \langle \Lambda^{p_1}\Phi_{j_1},B \Psi_{j_2}\rangle
    &\langle \Lambda^{p_2}\Phi_{j_2},B\Psi_{j_2}\rangle
    &\cdots &\langle \Lambda^{p_r}\Phi_{j_r},B\Psi_{j_2}\rangle \vspace{2mm}\\
    \vdots &\vdots &\ddots &\vdots \vspace{2mm}\\
    \langle \Lambda^{p_1}\Phi_{j_1},B\Psi_{j_r}\rangle
    &\langle \Lambda^{p_2}\Phi_{j_2},B\Psi_{j_r}\rangle
    &\cdots &\langle \Lambda^{p_r}\Phi_{j_r},B\Psi_{j_r}\rangle
    \end{array}\right|,
   \end{array} \label{eq:E}
\end{equation}
where $1\le i\le m$ and $l\ge 1$. Obviously,
each $E_{il}$ is a linear combination of the $F_{il}$'s, and hence
$\{E_{ik},E_{jl}\}=0$ holds for all $1\le i,j\le m$ and $k,l\ge 1$.
This means that the polynomial functions $E_{is}$, $1\le i\le m,$ $1\le s\le N$,
are also in involution in pair.

In order to 
show the functional independence of $E_{is},\ 1\le i\le m,\ 1\le s\le N$,
similar to the proof of Theorem \ref{maintheoremofintegralsofmotion2}, 
let $P_0$ be a point of $\R ^{2mN}$ satisfying
$\phi_{is}=\varepsilon$,
$1\le i\le m,$ $1\le s\le N$, where $\varepsilon$ is a
small constant. Then at this point $P_0$, we have 
\begin{equation}
   \frac{\partial E_{is_1}}{\partial\psi_{js_2}}
   =\varepsilon c_j^{i-1}\lambda_{s_2}^{s_1-1}\mu_{s_2}+\textrm{O}(\varepsilon^2),\ 1\le i,j\le m,\ 1\le s_1,s_2\le N.
\end{equation}
Hence a direct argument can give rise to 
\begin{equation}
   \begin{array}{rl}
   &\D\frac{\partial(E_{11},\cdots,E_{1N},
    E_{21},\cdots,E_{2N},\cdots,
    E_{m1},\cdots,E_{mN})}
   {\partial(\psi_{11},\cdots,\psi_{1N},
    \psi_{21},\cdots,\psi_{2N},\cdots,
    \psi_{m1},\cdots,\psi_{mN})}\\
   =&\D\varepsilon^{mN}\prod _{i=1}^N\mu _i
   \prod_{1\le i<j\le N}
   (\lambda_j-\lambda_i)^m
  \prod_{1\le i<j\le m}(c_j-c_i)^N
   +\textrm{O}(\varepsilon^{mN+1}).
   \vphantom{\vrule width 1pt height 28pt depth 0pt}
   \end{array}
\end{equation}
Therefore, if $c_1,c_2,\cdots, c_m$ are distinct,
the above Jacobian 
is not zero at $P_0$ when $\varepsilon\ne 0$ is small enough.
This implies that the functions $E_{is}$, $1\le i\le m,$ $ 1\le s\le N$,
are functionally independent over a dense open subset of $\R ^{2mN}$.

Let us sum up these results in the following theorem.

\begin{theorem}\label{thm:involutivepropertyofE_{is}}
All polynomial functions
in $\phi_{is}$ and $\psi_{is},$ $1\le i\le m$, $
1\le s\le N$: 
$E_{il}(c_1,\cdots,c_m)$, $1\le i\le m,$ $ l\ge 1$,
defined by (\ref{eq:E}),
are in involution in pair with respect to the Poisson bracket
(\ref{Poissonbracketofmorder}) for all constants $c_1,c_2,\cdots, c_m$.
Moreover, among them the polynomial functions 
$E_{is}(c_1,\cdots,c_m)$, $1\le i\le m,$ $ 1\le s\le N$,
are functionally independent over a dense open subset of 
$\R ^{2mN}$ for distinct constants $c_1,c_2,\cdots, c_m$.
\end{theorem}

Note that all polynomial functions $F_{il}$ are also linear combinations of 
the $E_{il}$'s.
The above theorem actually shows us an alternative 
to the involutive and functionally independent system 
of the polynomial functions $F_{is}$, $1\le i\le m,$ $1\le s\le N$.
The $E_{is}$'s have the compact form for the constants $c_1,c_2,\cdots, c_m$, and thus it is
more convenient to deal with them.  

\section{Liouville integrability and involutive solutions}
\setcounter{equation}{0}

Let us now turn to establish the Liouville integrability of the obtained constrained flows, 
and to present 
involutive solutions of the ${\cal N}$-wave interaction equations
%for the ${\cal N}$-wave interaction equations 
in both $1+1$ and $2+1$ dimensions.
The involutive system of the polynomial functions 
\[F_{is}=F_{is}(c_1,\cdots,c_m),\ 1\le i\le m,\ 1\le s\le N,\]
alternatively 
\[E_{is}=E_{is}(c_1,\cdots,c_m), \ 1\le i\le m,\ 1\le s\le N,\] 
will play an extremely important role in the following discussion. 

\subsection{Liouville integrability of the constrained flows}

For the 1+1 dimensional case, we have the matrix Lax operator as defined by 
(\ref{defofL^{(1)}}) and (\ref{defofL^{(1)}c}), i.e.,
\[ L^{(1)}(\la )=L^{(1)}(\la ;\gamma _1,\cdots,\gamma_n)=C_1(\gamma _1,\cdots,\gamma_n)
+D_1(\la ), \]
where $C_1$ and $D_1(\lambda )$ are given by (\ref{defofL^{(1)}c}).
Note that \[
\gamma_i\ne \gamma _j,\ 1\le i\ne j\le n.\]
According to Theorem \ref{maintheoremofintegralsofmotion1}
and Theorem \ref{maintheoremofintegralsofmotion2} for the case 
$ m=n$ and $ c_i=\gamma _i,\  1\le i\le n,$
we know that 
$F_{is}(\gamma _1,\cdots,\gamma _n),\  1\le i\le n,\ 1\le s
\le N,$
 defined by (\ref{defofF_{il}}), are functionally independent over a dense open subset 
of $\R ^{2nN}$ and in involution in pair
with respect to the Poisson bracket (\ref{eq:Poissonbracketin1+1case}), i.e.,
\[ \{f,g \}=
\sum_{i=1}^n\bigl( 
\langle \frac {\part f}{\part \Psi _{i}},B^{-1}\frac {\part g}{\part \Phi_{i}}\rangle
-\langle \frac {\part f}{\part \Phi _{i}},B^{-1}\frac {\part g}{\part \Psi_{i}}\rangle 
\bigr),\ f,g\in C^\infty (\R ^{2nN}).
\]

\begin{theorem} \label{Liouvilleintegrabilitytheoremofcf1+1}
Let $\gamma_1,\gamma _2,\cdots, \gamma _n$ be $n$ distinct numbers. Then
the spatial constrained flow (\ref{spatialconstrainedflowin1+1})
and the temporal constrained flow (\ref{temporalconstrainedflowin1+1})
of the $1+1$ dimensional $\cal N$-wave interaction equations 
(\ref{nWIEsin1+1}) are Liouville integrable Hamiltonian systems, which possess involutive and 
functionally independent integrals of motion
\[
F_{is}(\gamma_1,\cdots,\gamma_n), \ 1\le i\le n,\ 1\le s\le N,\] defined by 
(\ref{defofF_{il}}) in the case  
\[ m=n,\ c_i=\gamma _i,\ 1\le i\le n.\]
\end{theorem}

{\it Proof:} From the necessary Lax representations of 
the spatial constrained flow (\ref{spatialconstrainedflowin1+1})
and the temporal constrained flow (\ref{temporalconstrainedflowin1+1}):
\[(L^{(1)}(\la ))_x =[U(\tilde u ,\la ), L^{(1)}(\la )],\ 
(L^{(1)}(\la ))_{t_1} =[V^{(1)}(\tilde u,\la ), L^{(1)}(\la )],
\]
which are shown in Theorem \ref{thm:Hamiltoniansin1+1},
we can obtain \cite{MaS-PLA1994}
 \[
(L^{(1)}(\la ))^i)_x= 
[U(\tilde u,\la ),(L^{(1)}(\la ))^i],\ 
(L^{(1)}(\la ))^j)_{t_1}= 
[V^{(1)}(\tilde u,\la ),(L^{(1)}(\la ))^j],\ i,j\ge 1,
\]
and thus we have 
\[ \ba {l}
(\textrm{tr}(L^{(1)}(\la ))^i)_x=\textrm{tr}((L^{(1)}(\la ))^i)_x= 
\textrm{tr}[U(\tilde u,\la ),(L^{(1)}(\la ))^i]=0,\ i\ge 1,\vspace{2mm}\\
(\textrm{tr}(L^{(1)}(\la ))^j)_{t_1}=\textrm{tr}((L^{(1)}(\la ))^j)_{t_1}= 
\textrm{tr}[V^{(1)}(\tilde u,\la ),(L^{(1)}(\la ))^j]=0,\ j\ge 1.\ea 
\]
Therefore, ${\cal F}^{(k)}_\la (\gamma _1,\cdots,\gamma_n)$ are all generating functions of 
integrals of motion
of (\ref{spatialconstrainedflowin1+1})
and (\ref{temporalconstrainedflowin1+1}) in the light of the expression 
(\ref{Newtonf}) determined by Newton's identities.
It follows that 
$F_{is}(\gamma _1,\cdots,\gamma_n),$ $ 1\le i\le n,$ $ 1\le s\le N,$
are all integrals of motion of 
the spatial constrained flow (\ref{spatialconstrainedflowin1+1})
and the temporal constrained flow (\ref{temporalconstrainedflowin1+1}). Note that 
all constants $\gamma_1,\gamma_2,\cdots,\gamma_n$ are distinct. Therefore, Theorem \ref{maintheoremofintegralsofmotion1} and 
Theorem \ref{maintheoremofintegralsofmotion2} in the case of $m=n$ and $
c_i=\gamma _i,$ $1\le i\le n$, together with 
Theorem \ref{thm:Hamiltoniansin1+1},
show that the spatial constrained flow (\ref{spatialconstrainedflowin1+1})
and the temporal constrained flow (\ref{temporalconstrainedflowin1+1}) 
are Liouville integrable Hamiltonian systems,
which posses the involutive and functionally independent 
integrals of motion $F_{is}(\gamma_1,\cdots,\gamma_n)$, 
$1\le i\le n,\ 1\le s\le N$. The proof is finished.
$\vrule width 1mm height 3mm depth 0mm$

We remark that from the Lax representations shown in 
Theorem \ref{thm:Hamiltoniansin1+1},
we have 
\bea &&
 (\nu I_n-L^{(1)}(\la ))_x=[U({\tilde u},\la ) , \nu I_n-L^{(1)}(\la )],\nonumber \\
 && (\nu I_n-L^{(1)}(\la ))_{t_1}=[V^{(1)}({\tilde u},\la ) , \nu I_n-L^{(1)}(\la )]
\nonumber \eea 
for any constant $\nu $. It follows \cite{Tu-JPA1989}
that $\det (\nu I_n-L^{(1)}(\la )$ is 
a common generating function of 
integrals of motion
of the constrained flows (\ref{spatialconstrainedflowin1+1})
and (\ref{temporalconstrainedflowin1+1}), and thus so are 
${\cal F}_\la ^{(k)}(\gamma _1,\cdots,\gamma_n), $ $ 1\le k\le n$. This is an alternative proof
for showing that ${\cal F}_\la ^{(k)}(\gamma _1,\cdots,\gamma_n),$ $ 1\le k\le n$,
are the generating functions of integrals of motion
of (\ref{spatialconstrainedflowin1+1}) and (\ref{temporalconstrainedflowin1+1}).

For the $2+1$ dimensional case, a completely similar argument can give rise to 
the following theorem on the Liouville integrability of the constrained flows
(\ref{cf1of2+1case}), (\ref{cf2of2+1case}) and (\ref{cf3of2+1case})
of the $2+1$ dimensional 
$\cal N$-wave interaction equations (\ref{eq:eq_nw12}).

\begin{theorem} \label{Liouvilleintegrabilitytheoremofcf2+1}
Let $\delta _1,\cdots,\delta_n,\delta_{n+1}$ be $n+1$ distinct numbers. Then 
all three constrained flows (\ref{cf1of2+1case}), (\ref{cf2of2+1case}) and (\ref{cf3of2+1case})
of 
the $2+1$ dimensional $\cal N$-wave interaction equations (\ref{eq:eq_nw12})
are Liouville integrable Hamiltonian systems, which possess the involutive and
functionally independent integrals of motion
\[
F_{is}(\delta_1,\cdots,\delta_n,\delta_{n+1}), \  1\le i\le n+1, \ 
1\le s\le N,\]
 defined by 
(\ref{defofF_{il}}) in the case 
\[ m=n+1,\ c_i=\delta _i, \ 1\le i\le n+1.\]
\end{theorem}

\subsection{Involutive solutions of the $\cal N$-wave interaction equations }

We would like to show that the constrained flows provide 
involutive solutions to the $\cal N$-wave interaction equations in both $1+1$ and $2+1$ 
dimensions.
For the $1+1$ dimensional case, we have the following result.

\begin{theorem}\label{thm:involutivesolutionsin1+1}
If $\phi_{is}(x,t_1)$ and $\psi_{is}(x,t_1)$, $1\le i\le n$, $1\le s\le N$, 
solve the spatial constrained flow (\ref{spatialconstrainedflowin1+1}) 
and the temporal constrained flow (\ref{temporalconstrainedflowin1+1}) 
%of the $\cal N$-wave interaction equations in $1+1$ dimensions (\ref{nWIEsin1+1})
simultaneously,
then 
\be u_{ij}(x,t_1)=\frac {\al _i-\al _j}{\gamma _i -\gamma  _j}\langle\Phi_i(x,t_1) ,B\Psi_j(x,t_1)\rangle,
\ 1\le i\ne j\le n,
\label{espressionofu}\ee 
with $\Phi_i(x,t_1)$ and $\Psi_i(x,t_1)$ being given by 
\[ \Phi _i(x,t_1)=(\phi _{i1}(x,t_1),\cdots, \phi_{iN}(x,t_1))^T,\ 
\Psi_{i}(x,t_1)=(\psi_{i1}(x,t_1),\cdots,\psi_{iN}(x,t_1))^T,\
1\le i\le n,\]
solve the $1+1$ dimensional ${\cal N}$-wave interaction equations (\ref{nWIEsin1+1}). 
\end{theorem}

{\it Proof:}
Note that the $1+1$ dimensional ${\cal N}$-wave interaction equations
(\ref{nWIEsin1+1}) is the compatability condition of 
the spectral problem (\ref{spofnWIEsin1+1}) and the 
associated spectral problem (\ref{aspofmthnWIEsin1+1}) with $m=1$ or 
the adjoint spectral problem (\ref{aspofnWIEsin1+1}) and 
the adjoint associated spectral problem (\ref{aaspofmthnWIEsin1+1}) with $m=1$ for whatever potential $u$. Therefore, the $1+1$ dimensional ${\cal N}$-wave interaction equations
(\ref{nWIEsin1+1}) is also the compatability condition of 
the spatial constrained flow (\ref{spatialconstrainedflowin1+1}) 
and the temporal constrained flow (\ref{temporalconstrainedflowin1+1})
under the constraint (\ref{componentofgBscofnWIEsin1+1}).
Now $\phi_{is}(x,t_1)$ and $\psi_{is}(x,t_1)$, $1\le i\le n$, $1\le s\le N$, 
are assumed to solve (\ref{spatialconstrainedflowin1+1}) 
and (\ref{temporalconstrainedflowin1+1}) simultaneously,
and thus the potential defined by (\ref{espressionofu}) must satisfy 
the compatability condition of 
the spatial constrained flow (\ref{spatialconstrainedflowin1+1}) 
and the temporal constrained flow (\ref{temporalconstrainedflowin1+1}).
This means that the potential defined by (\ref{espressionofu}) must be a solution to 
the $1+1$ dimensional ${\cal N}$-wave interaction equations (\ref{nWIEsin1+1}). 
The proof is finished. 
$\vrule width 1mm height 3mm depth 0mm$

We remark that a direct computation can also show the above theorem. 
For the $2+1$ dimensional case, a similar deduction can give rise to the following theorem.

\begin{theorem}\label{thm:involutivesolutionsin2+1}
If $\phi_{is}(x,t)$ and $\psi_{is}(x,t)$, $1\le i\le n+1$, $1\le s\le N$, 
solve the constrained flows 
(\ref{cf1of2+1case}), (\ref{cf2of2+1case}) and (\ref{cf3of2+1case})
simultaneously, then 
\be \left\{ 
\begin{array} {l} p_{ij}(x,y,t)=\D \frac {J _i-J _j}{\delta _i -\delta 
  _j}\langle\Phi_i(x,y,t) ,B\Psi_j(x,y,t)\rangle,\ 1\le i\ne j\le n,
\vspace{2mm}\\
 q_{ij}(x,y,t)=
\D \frac {K _i-K _j}{\delta _i -\delta 
  _j}
\langle\Phi_i(x,y,t) ,B\Psi_j(x,y,t)\rangle,
\ 1\le i\ne j\le n,\end{array}\right.\label{isofNWIEs2+1}
\ee 
with $\Phi_i(x,t)$ and $\Psi_i(x,t)$ being given by 
\[ \Phi _i(x,t)=(\phi _{i1}(x,t),\cdots, \phi_{iN}(x,t))^T,\ 
\Psi_{i}(x,t)=(\psi_{i1}(x,t),\cdots,\psi_{iN}(x,t))^T,\
1\le i\le n+1,
\]
solve the $2+1$ dimensional ${\cal N}$-wave interaction equations
(\ref{eq:eq_nw12}).  
\end{theorem}

Also, one can find that 
\be f_i=\frac1{\delta _i-\delta _{n+1}}\langle\Phi_i,B\Psi_{n+1}\rangle
,\ g_i=\frac1{\delta _i-\delta _{n+1}}\langle\Phi_{n+1},B\Psi_i\rangle, \ 1\le i\le n
\label{isoff_iandg_i}
\ee 
provide a solution to the Lax system (\ref{eq:lp_nw12}) and the adjoint Lax system 
(\ref{eq:alp_nw12}) with the potentials given by (\ref{isofNWIEs2+1}).
What's more, (\ref{isofNWIEs2+1}) and (\ref{isoff_iandg_i}) 
automatically satisfy our first symmetry constraint (\ref{firstscofeq:eq_nw12}).

In the following theorem, the solutions 
given in Theorem~\ref{thm:involutivesolutionsin1+1}
and Theorem~\ref{thm:involutivesolutionsin2+1}
are shown to be involutive.

\begin{theorem}\label{thm:involutivepropertyofHamiltonians}
The Hamiltonians $H^x_1$ and $H^{t_1}_1$ of the constrained flows in $ 1+1$ dimensions,
defined by (\ref{H^x_1}) and (\ref{H^t_1}),
are the second-order polynomial functions of 
$E_{il}(\gamma_1,\cdots ,\gamma_n)$, 
$1\le i\le n$, 
$l=1,2,$
and thus they commute, i.e.,
\be \{H^x_1,H^{t_1}_1 \}=0,
%\sum_{i=1}^n (\langle \frac{\part H^x_1}{\part \Psi_i},B^{-1}\frac{\part H^t_1}{\part \Phi_i} %\rangle - \langle \frac{\part H^x_1}{\part \Phi_i},B^{-1}\frac{\part H^t_1}{\part \Psi_i} %\rangle )=0,
\ee 
where the Poisson bracket $\{\cdot,\cdot\}$ is defined by 
(\ref{eq:Poissonbracketin1+1case}).
The Hamiltonians $H^x_2$, $H^y_2$ and $H^t_2$ of the constrained flows in $ 2+1$ dimensions,
defined by (\ref{defofH_2^{x}}), (\ref{defofH_2^{y}}) and (\ref{defofH_2^{t}}),
are also the second-order polynomial functions of  
$E_{il}(\delta_1,\cdots,\delta_n,\delta_{n+1})$, $1\le i\le n+1$, $l=1,2,$
and thus they commute with each other, i.e.,
\be 
\{H^x_2,H^y_2 \}= \{H^x_2,H^t_2 \}=
\{H^y_2,H^t_2 \}=0 , \ee
where the Poisson bracket $\{\cdot,\cdot\}$ is defined by 
(\ref{eq:Poissonbracketin2+1case}).
\end{theorem}

{\it Proof:}
Directly from the explicit expression (\ref{eq:E}) of the $E_{is}$'s, we have 
\begin{eqnarray}
   E_{i1}&=&\D\sum_{j=1}^m c_j^{i-1}\langle\Phi_j,B\Psi_j\rangle,\ 1\le i\le m,
   \label{eq:E1}\\
   E_{i2}&=&\D\sum_{j=1}^m c_j^{i-1}\langle A\Phi_j,B\Psi_j\rangle
   \nonumber\\
   &&-\D\sum_{1\le j<k\le m}\frac{c_j^{i-1}-c_k^{i-1}}{c_j-c_k}\left(
    \langle\Phi_j,B\Psi_j\rangle\langle\Phi_k,B\Psi_k\rangle
    -\langle\Phi_j,B\Psi_k\rangle\langle\Phi_k,B\Psi_j\rangle\right)
   \nonumber\\
   &=&\D\sum_{j=1}^m c_j^{i-1}{\cal E}_j
    -\sum_{\scriptstyle j,k=1\atop\scriptstyle j\ne k}^m
    \frac{c_j^{i-1}}{c_j-c_k}
    \langle\Phi_j,B\Psi_j\rangle\langle\Phi_k,B\Psi_k\rangle,\ 1\le i\le m,
\label{eq:E2}
\end{eqnarray}
where the ${\cal E}_j$'s are defined as follows
\begin{equation}
   {\cal E}_j=\langle A\Phi_j,B\Psi_j\rangle
   +\sum_{\scriptstyle k=1\atop\scriptstyle k\ne j}^m
    \frac{1}{c_j-c_k}
    \langle\Phi_j,B\Psi_k\rangle\langle\Phi_k,B\Psi_j\rangle,\ 1\le j\le m.
\end{equation}
Now solving   
(\ref{eq:E1}) for $\langle\Phi_i,B\Psi_i\rangle$, $1\le i\le m$,
leads to
\begin{equation}
   \langle\Phi_i,B\Psi_i\rangle=
   \Bigl(\prod_{\scriptstyle r=1\atop\scriptstyle r\ne i}^m
   \frac 1 {c_i-c_r}\Bigr)
   \sum_{j=1}^m(-1)^{m-j}
   s_{m-j}(c_1,\cdots,c_{i-1},{\hat c}_i,c_{i+1},\cdots,c_m)E_{j1},
\ 1\le i\le m, \label{eq:Phi_iPsi_iintermsofE_1j}
\end{equation}
where the $s_j$'s are defined by (\ref{eq:s_j}) and 
${\hat c}_i$ means that $c_i$ does not appear. 
Therefore, 
each $\langle\Phi_i,B\Psi_i\rangle$ can be
expressed as a linear combination of $E_{i1}$, $1\le i\le m$. 
Similarly, solving (\ref{eq:E2}) for 
${\cal E}_j$, $ 1\le j\le m$,
leads to 
\begin{eqnarray}
   {\cal E}_i&=& \Bigl(\D\prod_{\scriptstyle r=1\atop\scriptstyle r\ne i}^m
   \frac 1{c_i-c_r}\Bigr)
   \sum_{j=1}^m(-1)^{m-j}
   s_{m-j}(c_1,\cdots,c_{i-1},{\hat c}_i,c_{i+1},\cdots,c_m)\times  \nonumber \\
   & &\D \Bigl(E_{j2}
   +\sum_{\scriptstyle k,l=1\atop\scriptstyle k\ne l}^m
   \frac{c_k^{j-1}}{c_k-c_l}
   \langle\Phi_k,B\Psi_k\rangle\langle\Phi_l,B\Psi_l\rangle\Bigr),\ 1\le i\le m. 
\end{eqnarray}
This expression together with (\ref{eq:Phi_iPsi_iintermsofE_1j})
implies that each ${\cal E}_j$ can be expressed as a linear combination of 
$E_{i1}$ and $E_{i2}$, $1\le i\le m$.

In the $1+1$ dimensional case,
%${\cal N}$-wave interaction equation,
we have $m=n$, $c_j=\gamma_j$, $1\le j\le n$. Hence
\begin{equation}
   {\cal E}_j=\D\langle A\Phi_j,B\Psi_j\rangle
   +\sum_{\scriptstyle k=1\atop\scriptstyle k\ne j}^n
   \frac 1{\gamma_j-\gamma_k}\langle\Phi_j,B\Psi_k\rangle\langle\Phi_k,B\Psi_j\rangle
,\ 1\le j\le n.\label{eq:calE_jin1+1}
\end{equation}
The Hamiltonians $H_1^x$ and $H_1^{t_1}$ 
in Theorem~\ref{thm:Hamiltoniansin1+1} can be easily expressed as
\begin{equation}
   H_1^x=-\sum_{k=1}^n\alpha_k{\cal E}_k,\quad
   H_1^{t_1}=-\sum_{k=1}^n\beta_k{\cal E}_k,
\end{equation}
where the ${\cal E}_k$'s are defined by (\ref{eq:calE_jin1+1}).

Likewise, in the $2+1$ dimensional case,
%${\cal N}$-wave interaction equation, 
we have $m=n+1$, $c_j=\delta_j, \ 1\le j\le n+1$. Hence
\begin{eqnarray} 
   {\cal E}_j&=&\D\langle A\Phi_j,B\Psi_j\rangle
   +\sum_{\scriptstyle k=1\atop\scriptstyle k\ne j}^n
   \frac 1{\delta_j-\delta_k}\langle\Phi_j,B\Psi_k\rangle\langle\Phi_k,B\Psi_j\rangle
\nonumber \\
   &&\D+\frac 1{\delta_j-\delta_{n+1}}\langle\Phi_j,B\Psi_{n+1}\rangle\langle\Phi_{n+1},B\Psi_j\rangle,\ 
1\le j\le n,\label{eq:calE_jin2+1} \\ 
{\cal E}_{n+1}&=&\D\langle A\Phi_{n+1},B\Psi_{n+1}\rangle
+\sum_{k=1}^n
   \frac 1{\delta_{n+1}-
\delta_k}\langle\Phi_{n+1},B\Psi_k\rangle\langle\Phi_k,B\Psi_{n+1}\rangle .
   \end{eqnarray}
The Hamiltonians $H_2^x$, $H_2^y$ and $H_2^t$ 
in Theorem~\ref{thm:Hamiltoniansin2+1} can be 
expressed as 
\begin{equation}
   H_2^x=-\sum_{k=1}^n{\cal E}_k,\quad
   H_2^y=-\sum_{k=1}^nJ_k{\cal E}_k,\quad
   H_2^t=-\sum_{k=1}^nK_k{\cal E}_k,
\end{equation}
where the ${\cal E}_k$'s are defined by (\ref{eq:calE_jin2+1}).

Therefore, 
$H_1^x$ and $H_1^{t_1}$ are linear combinations of 
$E_{il}(\gamma _1,\cdots, \gamma _n),\ 1\le i\le n, \ l=1,2$,
and 
$H_2^x$, $H_2^y$ and $H_2^t$ are 
linear combinations of 
$E_{il}(\delta _1,\cdots, \delta _n,\delta_{n+1}),\ 1\le i\le n+1, \ l=1,2$.
It follows from Theorem~\ref{thm:involutivepropertyofE_{is}} that 
$H_1^x$ and $H_1^{t_1}$ are in involution, and 
$H_2^x$, $H_2^y$ and $H_2^t$ are in involution in pair, too. 
The proof is finished. 
$\vrule width 1mm height 3mm depth 0mm$

We remark that a direct computation can also give a proof for 
the involutive property of the Hamiltonians 
of the constrained flows in both $1+1$ and $2+1$ dimensions.  
Only a new set of equalities
\[ \frac {a _j-a _i}{c _j-c _i}\frac 
{b _k-b _i}{c _k-c _i} -
\frac {a_k- a _i}{c_k-c _i}\frac 
{b _j-b _i}{c _j-c_i}+
\textrm{cycle}(i,j,k)=0,\ 1\le i,j,k\le n,
\] 
has to be utilized,
where $a_i$, $b_i$, and $c_i$, $1\le i\le n$, are arbitrary constants.
This just needs a direct check, too. 
However, the proof of Theorem~\ref{thm:involutivepropertyofHamiltonians} 
also gives rise to the explicit expressions for all Hamiltonians of the constrained flows
in both $1+1$ and $2+1$ dimensions, in terms of the integrals of motion 
$E_{is}$.

Now if we denote the Hamiltonian flows 
of the spatial constrained flow (\ref{spatialconstrainedflowin1+1})
and the temporal constrained flow (\ref{temporalconstrainedflowin1+1})
by $g_x^{H^x_1}$ and $g_t^{H^{t_1}_1}$ respectively,
then the above theorems present a kind of involutive solutions to 
the $1+1$ dimensional ${\cal N}$-wave interaction equations (\ref{nWIEsin1+1}):
\bea    
u_{ij}(x,t_1)&=&
 \frac {\al _i-\al _j}{\gamma  _i -\gamma  _j}\langle g_x^{H^x_1}g_t^{H^{t_1}_1}\Phi_{i0},
g_x^{H^x_1}g_t^{H^{t_1}_1}B\Psi_{j0}\rangle\nonumber \\
&=&\frac {\al _i-\al _j}{\gamma   _i -\gamma  _j}\langle g_t^{H^{t_1}_1}g_x^{H^x_1}\Phi_{i0},
g_t^{H^{t_1}_1}g_x^{H^x_1}B\Psi_{j0}\rangle,\ 1\le i\ne j\le n,
\label{involutivesolutionsof1+1case}
\eea  
where the initial values $\Phi_{i0}$ and $ \Psi_{i0}$ of $\Phi_{i}$ and $\Psi_i$
can be taken to be any arbitrary constant vectors of the Euclidean space $\R ^N$.
Similarly, if we denote the Hamiltonian flows of the constrained flows
(\ref{cf1of2+1case}), (\ref{cf2of2+1case}) and (\ref{cf3of2+1case})
by $g_x^{H^x_2}$, $g_y^{H^y_2}$ and $g_t^{H^t_2}$ respectively,
then the above theorems present a kind of involutive solutions to 
the $2+1$ dimensional ${\cal N}$-wave interaction equations (\ref{eq:eq_nw12}):
\bea    
p_{ij}(x,t)&=&
 \frac {J _i-J _j}{\delta  _i -\delta  _j}\langle g_x^{H^x_2}g_y^{H^y_2}g_t^{H^t_2}{\bar \Phi}_{i0},
g_x^{H^x_2}g_y^{H^y_2}g_t^{H^t_2}B{\bar \Psi}_{j0}\rangle\nonumber \\
&=&\frac {J _i-J _j}{\delta  _i -\delta  _j}
\langle g_y^{H^y_2}g_t^{H^t_2}g_x^{H^x_2}{\bar \Phi}_{i0},
g_y^{H^y_2}g_t^{H^t_2}g_x^{H^x_2}B{\bar \Psi}_{j0}\rangle \nonumber \\
&=&\frac {J _i-J _j}{\delta  _i -\delta  _j}
\langle g_t^{H^t_2}g_x^{H^x_2}g_y^{H^y_2}{\bar \Phi}_{i0},
g_t^{H^t_2}g_x^{H^x_2}g_y^{H^y_2}B{\bar \Psi}_{j0}\rangle \nonumber \\
&=& \cdots,\ 1\le i\ne j\le n
,\label{1stcinvolutivesolutionsof2+1case}
\\
q_{ij}(x,t)&=&
\frac {K _i-K _j}{\delta  _i -\delta  _j} 
\langle g_x^{H^x_2}g_y^{H^y_2}g_t^{H^t_2}{\bar \Phi}_{i0},
g_x^{H^x_2}g_y^{H^y_2}g_t^{H^t_2}B{\bar \Psi}_{j0}\rangle\nonumber \\
&=&\frac {K _i-K _j}{\delta  _i -\delta  _j} 
\langle g_y^{H^y_2}g_t^{H^t_2}g_x^{H^x_2}{\bar \Phi}_{i0},
g_y^{H^y_2}g_t^{H^t_2}g_x^{H^x_2}B{\bar \Psi}_{j0}\rangle \nonumber \\
&=&\frac {K _i-K _j}{\delta  _i -\delta  _j} 
\langle g_t^{H^t_2}g_x^{H^x_2}g_y^{H^y_2}{\bar \Phi}_{i0},
g_t^{H^t_2}g_x^{H^x_2}g_y^{H^y_2}B{\bar \Psi}_{j0}\rangle 
\nonumber \\
&=& \cdots,\ 1\le i\ne j\le n,
\label{2ndcinvolutivesolutionsof2+1case}
\eea  
where the initial values ${\bar \Phi}_{i0}$ and ${\bar \Psi}_{i0}$ 
of $\Phi_{i}$ and $\Psi_i$
can also be taken to be any arbitrary constant vectors of the Euclidean space $\R ^N$.

Note that all constrained flows in both $1+1$ and $2+1$ dimensions 
are Liouville integrable, and that 
%the arbitrariness of 
the initial values 
of $\Phi_{i}$ and $\Psi_i$, $1\le i\le n$, can be arbitrarily chosen.
Therefore, together with Theorem \ref{Liouvilleintegrabilitytheoremofcf1+1} and Theorem
\ref{Liouvilleintegrabilitytheoremofcf2+1},
the above involutive solutions also show us the richness of solutions and 
the integrability by quadratures 
for the ${\cal N}$-wave interaction equations in both $1+1$ and $2+1$ dimensions.
Of importance is of course that 
binary symmetry constraints decompose the ${\cal N}$-wave 
interaction equations in both $1+1$ and $2+1$ dimensions 
into finite-dimensional Liouville integrable Hamiltonian systems, and  
the resulting involutive solutions
present the B\"acklund transformations between 
the ${\cal N}$-wave interaction equations in both $1+1$ and $2+1$ dimensions
and these finite-dimensional Liouville integrable Hamiltonian systems. 

\section{Conclusions and remarks}
\setcounter{equation}{0}

We have introduced a class of special 
symmetry constraints, (\ref{gBscofnWIEsin1+1}) in the $1+1$ dimensional case, and 
(\ref{gB1scofnWIEsin2+1}) and (\ref{gB2scofnWIEsin2+1})
in the $2+1$ dimensional case, 
for the ${\cal N}$-wave interaction equations in both $1+1$ and 
$2+1$ dimensions. These symmetry constraints nonlinearize the $n\times n$
spectral problem and adjoint spectral problem, (\ref{Nreplicasofspandasp}) and 
(\ref{NreplicasofspofV^{(1)}}), and 
the $(n+1)\times (n+1)$
spectral problem and adjoint spectral problem, (\ref{eq:Phieq}) and 
(\ref{eq:Psieq}),
into finite-dimensional Liouville integrable Hamiltonian systems,
and decompose the 
${\cal N}$-wave interaction equations in both $1+1$ and $2+1$ dimensions
into these finite-dimensional Liouville integrable Hamiltonian systems.
A general involutive and functionally independent system of the polynomial functions
$F_{is}(c_1,\cdots,c_m),$ $1\le i\le m,$ $1\le s\le N$, or alternatively
$E_{is}(c_1,\cdots,c_m),$ $1\le i\le m,$ $1\le s\le N$,
associated with an arbitrarily higher-order matrix Lax operator,
was presented and used to show the Liouville integrability of the resulting 
constrained flows.
The nonlinear constraints on the potentials, resulting form the symmetry constraints,
also provide us with a class of B\"acklund transformations from    
the ${\cal N}$-wave interaction equations to 
the obtained finite-dimensional Liouville integrable systems.
The involutive solutions to the ${\cal N}$-wave interaction equations
are given through the constrained flows,
and thus the integrability by quadratures has been exhibited for 
the ${\cal N}$-wave interaction equations.
The special case with $\Gamma=W_0,$ i.e, $\diag (\gamma_1,\cdots,\gamma_n)=
\diag (\beta _1,\cdots,\beta _n)$
of two reductions of $n=3$ and $n=4$ in $1+1$ dimensions
presents all results established in \cite{WuG-JMP1999,ShiZ-JMP2000}.

We point out that for a more general matrix Lax operator $L=C+D$
with any constant matrix $C=(c_{ij})_{m\times m}$ and the matrix $D$ defined by (\ref{defofL(lambda)a}),
the ${\bf r}$-matrix formulation (\ref{rmatrixofNWIE}) still holds.
Therefore, an involutive system of polynomial functions can be generated, but 
we do not know what conditions on the matrix $C$ can ensure
the functional independence of that involutive system.
We are also curious about other examples of higher-order matrix Lax operators
which lead to involutive and functionally independent systems.
Our crucial techniques to present the involutive and functionally independent system 
$F_{is},\ 1\le i\le m,\ 1\le s\le N$, are the ${\bf r}$-matrix formulation, 
Newton's identities on elementary symmetric polynomials, and 
the determinant property of tensor products of matrices; and the whole process 
of their applications provides   
an efficient way to show the involutive property and the functional independence.  

Of course, one of the important results in binary nonlinearization 
is the integrability of soliton equations by quadratures,
%For the $\cal N$-wave interaction equations,
%the integrability shown 
%for the $\cal N$-wave interaction equations in 
%both $1+1$ and $2+1$ dimensions 
which implies that one can integrate 
%the ${\cal N}$-wave interaction equations 
soliton equations themselves by quadratures.
However, the potentials obtained by symmetry constraints 
can be proved to belong to a kind of finite-gap type solutions containing 
multi-soliton solutions,
and thus they may not present solutions to given initial value and/or boundary problems
of soliton equations. 
It is a challenging problem to establish a general theory of complete integrability for
nonlinear differential and differential-difference equations, 
which should state what mathematical properties the equations must possess 
so that their solutions to initial value and/or boundary problems
can also be determined by quadratures.

Symmetry constraints yield nonlinear constraints on potentials
of soliton equations, and put linear spectral problems (linear with respect to 
eigenfunctions) into nonlinear constrained flows (nonlinear again with
respect to eigenfunctions), which makes it more complicated to solve soliton equations. 
However, since spectral problems are overdetermined, one needs additional 
conditions (compatability conditions) to guarantee the existence of eigenfunctions of 
spectral problems. The symmetry property brings us 
the Liouville integrability for nonlinear constrained flows. 
Thus, symmetry constraints make up for the disadvantage of nonlinearization
in manipulating binary nonlinearization.
Of special interest in the study of symmetry constraints
are to create new classical integrable systems \cite{Ma-CSB1989},
which supplement the known class of integrable systems 
\cite{Perelomov-book1990}, and to expose the integrability by 
quadratures for soliton equations by using constrained flows \cite{MaF-book1996}. 

The idea of binary nonlinearization is quite similar to that of 
using adjoint symmetries to generate conservation laws for differential equations,
both Lagrangian and non-Lagrangian 
\cite{AncoB-PRL1997}. In binary nonlinearization, 
we adopt adjoint spectral problems to formulate Hamiltonian structures for 
constrained flows so that finite-dimensional Liouville integrable
systems result. Note that there exist also some special symmetry constraints which do not 
yield Hamiltonian structures with constant coefficient symplectic forms, including both canonical 
and nono-canonical ones, for constrained flows
\cite{MaL-PLA2000}. Therefore, 
it will be particularly interesting and 
important to classify symmetry constraints which 
exhibit Hamiltonian structures with constant and variable coefficient symplectic forms
for constrained flows.     

\vskip 8mm

\noindent {\bf Acknowledgments}

\noindent 
This work was supported by a grant from the Research Grants Council of Hong Kong 
Special Administrative Region, China (Project no. 9040466),
two grants from the City University of Hong Kong (Project nos. 7001041, 7001178),
the Chinese National Research
Project ``Nonlinear Science'', and the Doctoral Program Foundation
and the Foundation for University Key Teachers of the Ministry of
Education of China. The authors also acknowledge useful discussions with X. G. Geng,
Q. Y. Shi, Y. B. Zeng, R. G. Zhou and S. M. Zhu.

\newpage 
\appendix 
\renewcommand{\theequation}{\Alph{section}.\arabic{equation}}

%\small
\footnotesize

\section{Non-Lie symmetries}
\setcounter{equation}{0}

\begin{proposition} 
%Let $U_0=\textrm{diag}(\la _1,\al _2,\cdots ,\al _n)$ and $W_0=\textrm{diag}(\beta _1,\beta _2,\cdots ,\beta _n)$
%and $\mu_i,$ $1\le s\le N$, be nonzero constants.
If $\phi^{(s)}$ and $\psi^{(s)}$, $1\le s\le N$, satisfy 
(\ref{Nreplicasofspandasp}) and (\ref{NreplicasofspofV^{(1)}}), then 
the vector field 
\be 
Z_0=J\sum_{s=1}^N\mu _s \psi^{(s)T}\frac{\partial U(u,\lambda _s)}{\partial u}\phi^{(s)}
=\rho([U_0,\sum_{s=1}^N\mu_s \phi^{(s)}\psi^{(s)T}]),\ee
%$J$ and $\rho$ being defined by (\ref{defofJ}) and (\ref{eq:rho}) respectively,
is a symmetry of the 
$1+1$ dimensional ${\cal N}$-wave interaction equations (\ref{nWIEsin1+1}).
\end{proposition}

{\it Proof:}  
It is required to show that 
\be 
(\delta P ,\delta Q )=([U_0,\sum_{s=1}^N\mu_s \phi^{(s)}\psi^{(s)T}],
[W_0,\sum_{s=1}^N\mu_s \phi^{(s)}\psi^{(s)T}])
\label{2ndsymmetryofNWIEs1+1}
\ee 
satisfies the linearized system (\ref{lsofNWIEs1+1}).
By using (\ref{Nreplicasofspandasp}) and (\ref{NreplicasofspofV^{(1)}}),
we can first compute that 
\bea
(\sum_{s=1}^N\mu_s \phi^{(s)}\psi^{(s)T})_{t_1} &=&
 \sum_{s=1}^N\mu_s \phi^{(s)}_{t_1}
\psi^{(s)T}+\sum_{s=1}^N\mu_s \phi^{(s)}\psi^{(s)T}_{t_1}\nonumber\\
&=& \sum_{s=1}^N\mu_s V^{(1)}(u,\lambda _s)\phi^{(s)}
\psi^{(s)T}-\sum_{s=1}^N\mu_s \phi^{(s)}\psi^{(s)T}V^{(1)}(u,\lambda _s)\nonumber\\
&=& \sum_{s=1}^N\mu_s [V^{(1)}(u,\lambda _s),\phi^{(s)}
\psi^{(s)T}]\nonumber \\
 &=& \sum_{s=1}^N\lambda _s\mu_s [W_0,\phi^{(s)}
\psi^{(s)T}]+[W_1, \sum_{s=1}^N\mu_s \phi^{(s)}\psi^{(s)T}], \nonumber 
\eea
and similarly, we can have 
\[ (\sum_{s=1}^N\mu_s \phi^{(s)}\psi^{(s)T})_{x} =
\sum_{s=1}^N\lambda _s\mu_s [U_0,\phi^{(s)}
\psi^{(s)T}]+[U_1, \sum_{s=1}^N\mu_s \phi^{(s)}\psi^{(s)T}].
\]
Thus, noting the Jacobi identity, it follows that 
\bea
(\delta P) _{t_1}-(\delta Q) _x
&=&[U_0, (\sum_{s=1}^N\mu_s \phi^{(s)}\psi^{(s)T})_{t_1}]-[W_0,
(\sum_{s=1}^N\mu_s \phi^{(s)}\psi^{(s)T})_{x}]\nonumber \\
&=& \sum_{s=1}^N \lambda _s\mu_s 
([U_0,[W_0, \phi^{(s)}\psi^{(s)T} ]-[W_0,[U_0, \phi^{(s)}\psi^{(s)T}])
\nonumber \\
&& +[U_0,[W_1,\sum_{s=1}^N\mu_s \phi^{(s)}\psi^{(s)T}]-[W_0,[U_1,\sum_{s=1}^N\mu_s \phi^{(s)}\psi^{(s)T}]\nonumber \\
&=& [U_0,[W_1,\sum_{s=1}^N\mu_s \phi^{(s)}\psi^{(s)T}]-[W_0,[U_1,\sum_{s=1}^N\mu_s \phi^{(s)}\psi^{(s)T}] , \nonumber 
\eea
where $\delta P$ and $\delta Q$ are defined by (\ref{2ndsymmetryofNWIEs1+1}).
Then, again noting the Jacobi identity, we can have  
\bea 
&& (\delta P) _{t_1}-(\delta Q) _x+[U_1,\delta Q ]+[\delta P ,W_1]
\nonumber\\
 &=& [ \sum_{s=1}^N\mu_s \phi^{(s)}\psi^{(s)T},[U_0,W_1]]-
[\sum_{s=1}^N\mu_s \phi^{(s)}\psi^{(s)T},[W_0,U_1]]=0,\nonumber
\eea
in the last step of which we have used $[U_0,W_1]=[W_0,U_1]$. The proof is finished.
$\vrule width 1mm height 3mm depth 0mm$

All of the symmetries presented in this proposition are not Lie point, contact or 
B\"acklund symmetries, since they can not be 
written in terms of the potentials $u_{ij}$ and their spatial derivatives.  

\section{Formulas for computing ${{\cal F}}^{(k)}_\lambda $}
\setcounter{equation}{0}
\label{AppendixB}

Immediately from the expressions of ${{\cal F}}^{(k)}_\lambda $ in  
(\ref{eq:secondformulaforcalF^k_lambda}),
we can obtain the following more concrete 
formulas for computing ${{\cal F}}^{(k)}_\lambda $:
\bea  {{\cal F}}^{(1)}_\lambda &=&
 \sum_{i=1}^ m (c_i+ \stackrel{ii}Q_\lambda),
\nonumber\\
{{\cal F}}^{(2)}_\lambda &=&
 \sum_{1\leq i<j\leq m  }\left(c _ic _j+c _j\stackrel{ii}Q_\lambda
  +c _i\stackrel{jj}Q_\lambda+\left|\matrix{\stackrel{ii}Q_\lambda&\stackrel{ij}Q  _\lambda\vspace{2mm}
  \cr \stackrel{ji}Q_\lambda&\stackrel{jj}Q_\lambda\cr}\right|\right), 
\nonumber \\
{{\cal F}}^{(3)}_\lambda& =&
\sum_{1\leq i<j<k\leq m  }\left(c _ic _jc _k+c _ic _k\stackrel{jj}Q_\lambda+c _jc _k\stackrel{ii}Q_\lambda+c _ic _j\stackrel{kk}Q_\lambda\right)\nonumber \\
&& +\sum_{1\leq i<j<k\leq m  }\left(c _i\left|\matrix{\stackrel{jj}
Q_\lambda&\stackrel{jk}Q_\lambda
\vspace{2mm}
\cr \stackrel{kj}Q_\lambda&\stackrel{kk}Q_\lambda\cr}\right|+c _j\left|\matrix{\stackrel{ii}Q_\lambda&\stackrel{ik}Q_\lambda
\vspace{2mm}
\cr \stackrel{ki}Q_\lambda&\stackrel{kk}Q_\lambda \cr}\right|+c _k\left|\matrix{\stackrel{ii}Q_\lambda&\stackrel{ij}Q_\lambda
\vspace{2mm}
\cr \stackrel{ji}Q_\lambda&\stackrel{jj}Q_\lambda\cr}\right|+\left|\matrix
{\stackrel{ii}Q_\lambda&\stackrel{ij}Q_\lambda&\stackrel{ik}Q_\lambda
\vspace{2mm}
\cr \stackrel{ji}Q_\lambda&\stackrel{jj}Q_\lambda&\stackrel{jk}Q_\lambda
\vspace{2mm}
\cr \stackrel{ki}Q_\lambda&\stackrel{kj}Q_\lambda&\stackrel{kk}Q_\lambda\cr}\right|\right), 
\nonumber \\
&& \cdots \cdots\nonumber 
\\  
{{\cal F}}^{(k)}_\lambda&=&\sum_{1\leq j_1<j_2<\cdots<j_k\leq m  }\left(\prod^k_{p=1}c _{j_p}+\sum^k_{i=1}\prod^k_{\stackrel{p=1}{p\ne i}}c _{j_p}\stackrel{j_ij_i}Q_\lambda+\sum_{1\leq i_1<i_2\leq k}\prod^k_{\stackrel{p=1}{p\ne i_1,i_2}}c _{j_p}\left|\matrix{\stackrel{j_{i_1}j_{i_1}}Q_\lambda&\stackrel{j_{i_1}j_{i_2}}Q_\lambda\vspace{2mm}
\cr 
\stackrel{j_{i_2}j_{i_1}}Q_\lambda&\stackrel{j_{i_2}j_{i_2}}Q_\lambda\cr}\right|\right)
\nonumber \\
&&
+\sum_{1\leq j_1<j_2<\cdots<j_k\leq m  }\sum_{1\leq i_1<i_2<i_3\leq k}\prod^k_{\stackrel{p=1}{p\ne i_1,i_2,i_3}}c _{j_p}\left|\matrix{\stackrel{j_{i_1}j_{i_1}}
Q_\lambda&\stackrel{j_{i_1}j_{i_2}}Q_\lambda&
\stackrel{j_{i_1}j_{i_3}}Q_\lambda\vspace{2mm}
\cr \stackrel{j_{i_2}j_{i_1}}Q_\lambda&\stackrel{j_{i_2}j_{_2}}
Q_\lambda&\stackrel{j_{i_2}j_{i_3}}Q_\lambda\vspace{2mm}
\cr \stackrel{j_{i_3}j_{i_1}}Q_\lambda&\stackrel{j_{i_3}j_{i_2}}Q_\lambda&\stackrel{j_{i_3}j_{i_3}}Q_\lambda\cr}\right|
\nonumber \\
&&  +\cdots+\sum_{1\leq j_1<j_2<\cdots<j_k\leq m  }\left|
\matrix{\stackrel{j_1j_1}Q_\lambda&\stackrel{j_1j_2}Q_\lambda&\cdots&\stackrel{j_1j_k}Q_\lambda\vspace{2mm}
\cr \stackrel{j_2j_1}Q_\lambda&\stackrel{j_2j_2}Q_\lambda&\cdots&\stackrel{j_2j_k}
Q_\lambda\vspace{2mm}
\cr \vdots&\vdots&\ddots&\vdots\vspace{2mm}
\cr \stackrel{j_kj_1}Q_\lambda&\stackrel{j_kj_2}Q_\lambda&\cdots&\stackrel{j_kj_k}Q_\lambda\cr}\right|,\nonumber \\
&& \cdots \cdots\nonumber 
\\ 
{{\cal F}}^{(m  )}_\lambda&=&\prod^m  _{p=1}c _{p}+\sum^m  _{i=1}\prod^m  _{\stackrel{p=1}{p\ne i}}c _{p}\stackrel{ii}Q_\lambda+\sum_{1\leq i<j\leq m  }\prod^m  _{\stackrel{p=1}{p\ne i,j}}c _{p}\left|\matrix{\stackrel{ii}Q_\lambda&\stackrel{ij}Q_\lambda\vspace{2mm}
\cr \stackrel{ji}Q_\lambda&\stackrel{jj}Q_\lambda\cr}\right|
\nonumber \\
&& +\sum_{1\leq i<j<k\leq m  }\prod^k_{\stackrel{p=1}{p\ne i,j,k}}c _{p}\left|\matrix{\stackrel{ii}Q_\lambda&\stackrel{ij}Q_\lambda&\stackrel{ik}Q_\lambda
\vspace{2mm}
\cr \stackrel{ji}Q_\lambda&\stackrel{jj}Q_\lambda&\stackrel{jk}Q_\lambda
\vspace{2mm}
\cr \stackrel{ki}Q_\lambda&\stackrel{kj}Q_\lambda&\stackrel{kk}Q_\lambda\cr}\right|
+\cdots+\left|\matrix{\stackrel{11}Q_\lambda&\stackrel{12}
Q_\lambda&\cdots&\stackrel{1m  }Q_\lambda\vspace{2mm}
\cr \stackrel{21}Q_\lambda&\stackrel{22}Q_\lambda&\cdots&\stackrel{2r }Q_\lambda
\vspace{2mm}
\cr \vdots&\vdots&\ddots&\vdots
\vspace{2mm}
\cr \stackrel{m  1}Q_\lambda&\stackrel{r 2}Q_\lambda&\cdots&\stackrel{m  m  }Q_\lambda\cr}\right|.
\nonumber
% \label{defofcalF^{(m)}_lambda}
\eea

\section{The determinant of $\Omega _m$}\label{app:detofOmega_m}
\setcounter{equation}{0}

The following proposition has been used 
while showing the functional independence of 
the polynomial functions
$F_{is}(c_1,\cdots,c_m),\ 1\le i\le m,\ 1\le s
\le N$, which  
is of interest itself.

\begin{proposition}
\label{lemma:anewNthorderdet}
Let $m\ge 2$, and $c_1,c_2,\cdots,c_m$ be constants. Then 
\be 
\det (\Omega _m)=
\left |
\ba {cccccc} 1 \ &\D  \sum_{i=2}^m c_i & \D \sum_{2\le i<j\le m}c_ic_j &\D \sum_{2\le i<j<k\le m}c_ic_jc_k &
\cdots & 
\D \prod _{i=2}^mc_i \vspace{2mm}\\
 1 \ & \D \sum_{i=1\atop i\ne 2}^m c_i &\D  \sum_{1\le i<j\le m\atop i,j\ne 2}c_ic_j &\D \sum_{1\le i<j<k\le m\atop i,j,k\ne 2}c_ic_jc_k &
\cdots & \D 
\prod _{i=1\atop i\ne 2}^mc_i  \vspace{2mm}\\
\vdots \ &\vdots &\vdots & \vdots &\ddots & \vdots \vspace{2mm}\\
1 \ &\D  \sum_{i=1}^{m-1} c_i &\D  \sum_{1\le i<j\le m-1}c_ic_j &\D \sum_{1\le i<j<k\le m-1}c_ic_jc_k &\cdots & 
\D \prod _{i=1}^{m-1}c_i 
 \ea 
\right|
 = \prod_{1\le i<j\le m}(c_i-c_j).
\label{detofDelta}
\ee 
\end{proposition}

{\it Proof:} We prove this proposition by the principle of 
mathematical induction.
It is obvious that 
(\ref{detofDelta}) is true when $m=2$. 
Suppose that (\ref{detofDelta}) is true when $m=l$. 
Let us verify that (\ref{detofDelta}) is also true when $m=l+1$. 
Note that 
\bea &&
\sum_{1\le i_1<i_2<\cdots <i_k\le l+1\atop i_1,i_2,\cdots ,i_k\ne 
j}c_{i_1}c_{i_2}\cdots {c_{i_k}}-
\sum_{1\le i_1<i_2<\cdots <i_k\le l+1\atop i_1,i_2,\cdots ,i_k\ne 
i}c_{i_1}c_{i_2}\cdots {c_{i_k}}\nonumber \vspace{2mm}
\\
&&=
(c_i-c_j)
\sum_{1\le i_1<i_2<\cdots <i_{k-1}\le l+1\atop i_1,i_2,\cdots ,i_{k-1}\ne 
i,j}c_{i_1}c_{i_2}\cdots {c_{i_{k-1}}},
\ 1\le i,j\le l+1,\ 1\le k\le l.   \nonumber \eea 
For each $2\le j\le l+1$,
we subtract 
\[ \D  \sum_{2\le i_1<i_2<\cdots <i_ {j-1}\le l+1 }^{l+1} c_{i_1}
c_{i_2}\cdots c_{i_{j-1}}
\times \textrm{the first column of det($\Omega  _{l+1}$)} 
\]
from the $j$th column of det($\Omega  _{l+1}$), and then we have 
\bea &&
\textrm{det}(\Omega _{l+1}) \nonumber \\ && \quad \nonumber \\ &=&
\left |
\ba {cccccc} 
1& 0& 0 &0&\cdots & 0\\
\\
1 & c_1-c_2& (c_1-c_2)\D \sum_{i=3}^{l+1} c_i & (c_1-c_2)\D \sum_{3\le i<j\le l+1}c_ic_j 
 & \cdots & 
(c_1-c_2)\D \prod _{i=3}^{l+1}c_i \vspace{2mm}\\
1 & c_1-c_3& (c_1-c_3)\D \sum_{i=2\atop i\ne 3 }^{l+1} c_i & (c_1-c_3)\D 
\sum_{2\le i<j\le l+1
\atop i,j\ne 3}c_ic_j 
 & \cdots & 
(c_1-c_3)\D \prod _{i=2\atop i\ne 3}^{l+1}c_i \vspace{2mm}\\
\vdots &\vdots &\vdots &\vdots &\ddots & \vdots \vspace{2mm}\\
1 & c_1-c_{l+1}& (c_1-c_{l+1})\D \sum_{i=2}^{l} c_i &(c_1-c_{l+1}) 
\D \sum_{2\le i<j\le l}c_ic_j 
&\cdots & (c_1-c_{l+1})
\D \prod _{i=2}^{l}c_i 
 \ea 
\right| \nonumber \\
&=& \D \prod_{j=2}^{l+1}(c_1-c_j)
\left |
\ba {ccccc} 
1\  & \D \sum_{i=3}^{l+1} c_i & \D \sum_{3\le i<j\le l+1}c_ic_j 
 & \cdots & 
\D \prod _{i=3}^{l+1}c_i \vspace{2mm}\\
1 \ & \D \sum_{i=2\atop i\ne 3 }^{l+1} c_i & \D \sum_{2\le i<j\le l+1
\atop i,j\ne 3}c_ic_j 
 & \cdots & 
\D \prod _{i=2\atop i\ne 3}^{l+1}c_i \vspace{2mm}\\
\vdots \ &\vdots &\vdots &\ddots & \vdots \vspace{2mm}\\
1 \ & \D \sum_{i=2}^{l} c_i & \D \sum_{2\le i<j\le l}c_ic_j 
&\cdots & \D \prod _{i=2}^{l}c_i 
 \ea 
\right| = \D \prod_{1\le i<j\le l+1}(c_i-c_j),
\nonumber \eea 
in the last step of which we have used the inductive assumption.
This means that (\ref{detofDelta}) is also true when $m=l+1$, i.e.,
the inductive step is satisfied.
Therefore, the formula (\ref{detofDelta}) is always true by 
the principle of mathematical induction. The proof is finished. 
$\vrule width 1mm height 3mm depth 0mm$

\section{Two identities on symmetric polynomials}
\setcounter{equation}{0}
\label{appx:twoidentitiesforsp}

Let the $s_j$'s be symmetric polynomials defined by (\ref{eq:s_j}). 

\begin{proposition}\label{prop:sum1}
For any integers $r$ and $i$ with $i\ge r\ge 1$, and any
numbers $c_1,\cdots, c_r$, we have 
\begin{equation}
   \sum_{j=0}^{i-r}(-1)^j\sj{j}{r}\llsum{r}{i-r-j}
   =\left\{\begin{array}{ll} 1,\quad &\hbox{\rm if}\ i=r,\vspace{2mm} \\
    0,\quad &\hbox{\rm if}\ i>r.\end{array}\right.
   \label{eq:sum1}
\end{equation}
\end{proposition}

{\it Proof:}
Use the principle of mathematical induction on $r$. When $r=1$ and $i=1$, the left-hand side of (\ref{eq:sum1}) is $1$.
When $r=1$ and $i>1$, the left-hand side of (\ref{eq:sum1}) is $0$. Hence (\ref{eq:sum1}) holds
when $r=1$.

Now suppose that (\ref{eq:sum1}) holds when $r=k$, i.e.,
\begin{equation}
   \sum_{j=0}^{i-k}(-1)^j\sj{j}{k}\llsum{k}{i-k-j}
   =\left\{\begin{array}{ll} 1,\quad &\hbox{\rm if}\ i=k,\vspace{2mm}\\
    0,\quad &\hbox{\rm if}\ i>k.\end{array}\right.
\label{assumptionofr=k}
\end{equation}
Then, when $r=k+1$, the left-hand side of (\ref{eq:sum1}) is
\begin{equation}
   \sum_{j=0}^{i-k-1}(-1)^j\sj{j}{k+1}\llsum{k+1}{i-k-j-1}.
\end{equation}
By using (\ref{eq:sdecomp}), it equals to
\begin{eqnarray}
   %\begin{array}{rl}
   &&\displaystyle \sum_{j=0}^{i-k-1}(-1)^j
   \sum_{l_{r+1}=0}^{i-k-j-1}c_{k+1}^{l_{k+1}+1}
   \sj{j-1}{k}\llsum{k}{i-k-j-1-l_{k+1}}\nonumber \\
   &&\displaystyle +\sum_{j=0}^{i-k-1}(-1)^j
   \sum_{l_{r+1}=0}^{i-k-j-1}c_{k+1}^{l_{k+1}}
   \sj{j}{k}\llsum{k}{i-k-j-1-l_{k+1}}\nonumber  \\
   &=&\displaystyle \sum_{l_{r+1}=0}^{i-k-1}c_{k+1}^{l_{k+1}+1}
   \sum_{j=0}^{i-k-l_{k+1}-2}(-1)^{j+1}
   \sj{j}{k}\llsum{k}{i-k-j-l_{k+1}-2}\nonumber \\
   &&\displaystyle +\sum_{l_{r+1}=0}^{i-k-1}c_{k+1}^{l_{k+1}}
   \sum_{j=0}^{i-k-l_{k+1}-1}(-1)^j
   \sj{j}{k}\llsum{k}{i-k-j-l_{k+1}-1},
   %\end{array}
   \label{eq:sum1_ss}
\end{eqnarray}
where an empty sum is understood to be zero.
% if the lower bound is greater than the upper bound.

When $i=k+1$, it is easy to see that (\ref{eq:sum1_ss}) equals to $1$. If $i>k+1$,
then by (\ref{assumptionofr=k}), the first sum equals to
\begin{equation}
   \bigl.-c_{k+1}^{l_{k+1}+1}\bigr|_{l_{k+1}=i-k-2}=-c_{k+1}^{i-k-1},
\end{equation}
and again, by (\ref{assumptionofr=k}), the second sum equals to
\begin{equation}
   \bigl.c_{k+1}^{l_{k+1}}\bigr|_{l_{k+1}=i-k-1}=c_{k+1}^{i-k-1}.
\end{equation}
Hence (\ref{eq:sum1_ss}) equals to $0$ if $i>k+1$,
which implies that (\ref{eq:sum1}) holds when $r=k+1$.
Therefore, (\ref{eq:sum1}) always holds by the principle of mathematical induction.
 The proposition is proved.
$\vrule width 1mm height 3mm depth 0mm$

\begin{proposition}\label{prop:sum2}
For any integers $m$, $r$, $i$ with $i\ge r+1\ge 2$, $m$
numbers $c_1,\cdots,c_m$, and $r$ integers
$j_1,\cdots,j_r$ with $1\le j_1<\cdots<j_r\le m$, we have
\begin{equation}
   \sum_{j=0}^{i-r}(-1)^{i-r-j}\sj{j}{m}
   \sum_{\scriptstyle l_1+\cdots+l_r=i-r-j\atop
    \scriptstyle l_1,\cdots,l_r\ge 0}
   c_{j_1}^{l_1}\cdots c_{j_r}^{l_r}
   =\sum_{\scriptstyle 1\le\rho_1<\cdots<\rho_{i-r}\le m\atop
    \scriptstyle \rho_\alpha\ne j_\beta\hbox{\scriptrm \ for all }\alpha,\beta}
   c_{\rho_1}\cdots c_{\rho_{i-r}}.
   \label{eq:sum2}
\end{equation}
\end{proposition}

{\it Proof:}
Without loss of generality, suppose that $j_i=i$ when $i=1,\cdots,r$, since 
each $s_j(c_1,\cdots,c_m)$ is symmetric with respect to 
$c_1,\cdots,c_m$. Then, 
(\ref{eq:sum2}) becomes
\begin{equation}
   \sum_{j=0}^{i-r}(-1)^{i-r-j}\sj{j}{m}\llsum{r}{i-r-j}
   =\sum_{r+1\le\rho_1<\cdots<\rho_{i-r}\le m}
   c_{\rho_1}\cdots c_{\rho_{i-r}}.
   \label{eq:sum2_1}
\end{equation}
Obviously, for any fixed $j$ with $r+1\le j\le m$, 
both sides of (\ref{eq:sum2_1}) are linear
with respect to $c_j$.

We use the principle of mathematical induction on $i$ to prove (\ref{eq:sum2_1}).
When $i=r+1$, both sides of (\ref{eq:sum2_1}) equal to $c_{r+1}+\cdots+c_m$.

Suppose that (\ref{eq:sum2_1}) holds when $i=k$ $(k> r)$. Then, when
$i=k+1$, the left-hand side of (\ref{eq:sum2_1}) reads as 
\begin{eqnarray}
%   \begin{array}{rl}
   R&:=&\displaystyle\sum_{j=0}^{k+1-r}(-1)^{k+1-r-j}\sj{j}{m}\llsum{r}{k+1-r-j}
\nonumber \\
   &=&\displaystyle\sum_{j=-1}^{k-r}(-1)^{k-r-j}\sj{j+1}{m}\llsum{r}{k-r-j}.
\label{eq:sum2_2}
   %\end{array}
\end{eqnarray}
Then by (\ref{eq:sdecomp}), we have 
\begin{equation}
   \frac{\partial R}{\partial c_m}
   =\sum_{j=0}^{k-r}(-1)^{k-r-j}\sj{j}{m-1}\llsum{r}{k-r-j}.
\end{equation}
By the inductive assumption, it becomes
\begin{equation}
   \frac{\partial R}{\partial c_m}
   =\sum_{r+1\le\rho_1<\cdots<\rho_{k-r}\le m-1}
   c_{\rho_1}\cdots c_{\rho_{k-r}}.
\end{equation}
Hence we obtain
\begin{equation}
   R=\sum_{r+1\le\rho_1<\cdots<\rho_{k-r}\le m-1}
   c_{\rho_1}\cdots c_{\rho_{k-r}}c_m+R_1(c_1,\cdots,c_{m-1}),
\end{equation}
where $R_1$ is a polynomial. Since $R$ is symmetric with respect to $c_{r+1},\cdots,c_m$, we have 
\begin{equation}
   R=\sum_{r+1\le\rho_1<\cdots<\rho_{k+1-r}\le m}
   c_{\rho_1}\cdots c_{\rho_{k+1-r}}+R_0(c_1,\cdots,c_r),
\end{equation}
where by setting $c_{r+1}=\cdots =c_m=0$ in (\ref{eq:sum2_2}), $R_0$ 
is determined to be 
\begin{equation}
   R_0(c_1,\cdots,c_r)=\sum_{j=0}^{k+1-r}(-1)^{k+1-r-j}\sj{j}{r}
   \llsum{r}{k+1-r-j}.
\end{equation}
By Proposition~\ref{prop:sum1}, $R_0=0$ since $k+1=i>r$. Hence
\begin{equation}
   R=\sum_{r+1\le\rho_1<\cdots<\rho_{k+1-r}\le m}
   c_{\rho_1}\cdots c_{\rho_{k+1-r}},
\end{equation}
which implies that (\ref{eq:sum2_1}) holds when $i=k+1$.
Therefore, (\ref{eq:sum2_1}) holds for all $i>r$
by the principle of mathematical induction.
The proof is completed.
$\vrule width 1mm height 3mm depth 0mm$

The identity (\ref{eq:sum2}) is needed in
presenting an alternative involutive system $E_{is}$'s to the $F_{is}$'s
in the subsection~\ref{subsec:alternativeisE_{is}}.

\end{document}